\documentclass[11pt]{article} 
\usepackage[left=2.5cm, right=2.5cm, top=1.785cm, bottom=2.0cm]{geometry}

\usepackage[utf8]{inputenc}
\usepackage{amsmath}
\usepackage{amsfonts}
\usepackage{amssymb}
\usepackage{mathtools}
\usepackage{listings}
\usepackage{mathrsfs}
\usepackage{times}
\usepackage{float}
\usepackage{color}
\usepackage{setspace}
\usepackage{color,soul}

\usepackage{amsmath,amsfonts,amsthm,eucal}
\usepackage{enumerate}
\usepackage{color}
\usepackage{tabularx}
\usepackage{siunitx}
\usepackage{empheq}
\usepackage[tableposition=below]{caption}

\usepackage[
pdfstartview=XYZ,
bookmarks=true,
colorlinks=true,
linkcolor=blue,
urlcolor=blue,
citecolor=blue,
pdftex,
bookmarks=true,
linktocpage=true,   
hyperindex=true
]{hyperref}

\usepackage{lipsum}
\usepackage[FIGBOTCAP,TABTOPCAP,tight]{subfigure}

\usepackage[pdftex]{graphicx}
\usepackage{epstopdf}
\usepackage{booktabs} 
\usepackage{tikz,mathpazo}
\usetikzlibrary{shapes.geometric, arrows}
\usepackage{caption}

\usepackage{longtable}
\usepackage{adjustbox}
\usepackage{framed}

\usepackage[colorinlistoftodos]{todonotes}

\usepackage{amsmath}

\usepackage{nicefrac}
\usepackage{multirow}
\usepackage{hhline}
\usepackage{lineno}
\usepackage{import}
\usepackage{epstopdf}
\usepackage{subcaption}
\usepackage{mathtools}
\usepackage{transparent}

\usepackage{cancel}
\usepackage{empheq}
\usepackage{booktabs}
\usepackage{cleveref}

\crefname{equation}{Eq.}{Eqs.}
\usepackage{soul}
\usepackage{courier}

\definecolor{tbf}{RGB}{255,0,0} 
\definecolor{txue}{RGB}{0,0,255}

\usepackage{authblk}
\usepackage{listings}
\usepackage{titlesec}
\usepackage[numbers,sort&compress]{natbib}

\newcommand\keywords[1]{\textbf{Keywords}: #1}

\title{\Large DiffLiB: High-fidelity differentiable modeling of lithium-ion batteries and efficient gradient-based parameter identification} 

\begin{document}

\author[1]{\normalsize Weipeng Xu}
\author[2]{\normalsize Kaiqi Yang}
\author[2]{\normalsize Yuzhi Zhang}
\author[3]{\normalsize Shichao Sun}
\author[3]{\normalsize Sheng Mao}
\author[1]{\normalsize Tianju Xue
\footnote{\textit{cetxue@ust.hk}  (corresponding author)}}
\affil[1]{\footnotesize Department of Civil and Environmental Engineering, Hong Kong University of Science and Technology, HKSAR, China}
\affil[2]{\footnotesize DP Technology, Beijing, China}
\affil[3]{\footnotesize Department of Mechanics and Engineering Science, College of Engineering, Peking University, Beijing, China}

\date{}
\maketitle

\vspace{-30pt}

\begin{abstract}
The physics-based Doyle-Fuller-Newman (DFN) model, widely adopted for its precise electrochemical modeling, stands out among various simulation models of lithium-ion batteries (LIBs).
Although the DFN model is powerful in forward predictive analysis, the inverse identification of its model parameters has remained a long-standing challenge.
The numerous unknown parameters associated with the nonlinear, time-dependent, and multi-scale DFN model are extremely difficult to be determined accurately and efficiently, hindering the practical use of such battery simulation models in industrial applications.
To tackle this challenge, we introduce \texttt{DiffLiB}, a high-fidelity finite-element-based LIB simulation framework, equipped with advanced differentiable programming techniques so that efficient gradient-based inverse parameter identification is enabled.
Customized automatic differentiation rules are defined by identifying the VJP (vector-Jacobian product) structure in the chain rule and implemented using adjoint-based implicit differentiation methods. 
Four numerical examples, including both 2D and 3D forward predictions and inverse parameter identification, are presented to validate the accuracy and computational efficiency of \texttt{DiffLiB}. 
Benchmarking against \texttt{COMSOL} demonstrates excellent agreement in forward predictions, with terminal voltage discrepancies maintaining a root-mean-square error (RMSE) below $2\rm~mV$ across all test conditions. 
In parameter identification tasks using experimentally measured voltage data,
the proposed gradient-based optimization scheme achieves superior computational performance, with $96\%$ fewer forward predictions and $72\%$ less computational time compared with gradient-free approaches. 
These results demonstrate that \texttt{DiffLiB} is a versatile and powerful computational framework for the development of advanced LIBs.
\end{abstract}


\section{Introduction}
\label{intro}
In recent years, lithium-ion batteries (LIBs) have emerged as one of the most critical modern energy storage technologies. Compared to other rechargeable batteries, LIBs offer the advantages of high energy density and electrochemical stability, enabling their widespread applications in portable electronic devices, electric vehicles, and energy storage systems~\cite{goodenough2013li}. With the growing demand for sustainable development, LIBs have become essential components for promoting the low-carbon economy and achieving carbon neutrality~\cite{lin2023carbon}.

The performance of LIBs is governed by internal electrochemical reactions such as ionic transport and interface reactions. Understanding these mechanisms is crucial for the design optimization and safety management of LIBs. While experimental methods such as cycling~\cite{li2001studies} and pulse charging~\cite{li2001effects} tests can provide valuable insights, they are often limited by high costs, time-consuming processes, and the inability to fully capture internal states. In contrast, numerical simulation offers a more efficient and economical approach to analyze the internal states, significantly shortening the research and development cycle of LIBs.

Numerous battery simulation models have been developed to address analysis problems in different scenarios. According to the modeling principles, these models can be classified into empirical models and physical models. Empirical models, such as equivalent circuit methods (ECMs)~\cite{hu2012comparative} and data-driven approaches~\cite{dawson2018data}, attempt to establish the input-output mapping relationship with experimental measurements. ECMs have been widely adopted in battery management systems (BMS) for electric vehicles due to their computational efficiency and simplicity. However, they lack the capability to describe internal states. Meanwhile, data-driven models suffer from generalization, and their performance heavily depends on the quality of training data~\cite{ali2024comparison}. On the contrary, the physical models are built on electrochemical mechanisms and can naturally capture internal states within LIBs. The pioneering work of physical models is the Doyle-Fuller-Newman (DFN) model based on Newman's porous electrode theory~\cite{doyle1993modeling,fuller1994simulation, newman2021electrochemical}. Subsequent studies introduced extensions accounting for additional physical phenomena, such as thermal-mechanical effects~\cite{kulathu2024three} and aging mechanisms~\cite{lamorgese2018electrochemical}. Reduced-order models (ROM) like the Single Particle Model (SPM)~\cite{ning2004cycle,han2015simplification} are also developed to enhance computational efficiency. The high precision and excellent adaptability of the DFN model establish it as the most widely used battery simulation model, as well as the first choice for most battery simulation packages, such as LIONSIMBA~\cite{torchio2016lionsimba}, PyBaMM~\cite{Sulzer2021}, cideMOD~\cite{aylagas2022cidemod}, and JuBat~\cite{ai2024jubat}.

Despite the great success of the DFN model in forward predictive analysis, efficient inverse identification of numerous model parameters is a long-standing challenge.
One established method is the destructive physical analysis (DPA)~\cite{chen2020development}, which involves disassembling battery cells and measuring the parameters with specialized instruments. 
Although DPA provides detailed insights into battery properties, this process is usually invasive and time-consuming. 
As a promising alternative for DPA, numerical optimization~\cite{nocedal1999numerical} provides a non-destructive way for parameter identification. The model parameters are iteratively updated to minimize the discrepancy between model predictions and experimental measurements. 
Owing to its efficiency and non-destructive nature, this approach has emerged as a research focus in recent years.

Depending on whether gradient information is utilized during the optimization process, parameter identification methods can be divided into two categories: gradient-free and gradient-based identification methods. Population-based algorithms, such as genetic algorithms (GA)~\cite{mitchell1998introduction} and particle swarm optimization (PSO)~\cite{kennedy1995particle}, dominate gradient-free identification methods. These approaches iteratively evolve an initial population through stochastic operators (e.g., selection, crossover, mutation, or velocity updates) to converge toward optimal parameter combinations. Forman et al.~\cite{forman2012genetic} conducted voltage-current cycling experiments and applied the GA to identify the complete parameters of the full-order DFN model. Zhang et al.~\cite{zhang2014multi} employed the multi-objective NSGA-II algorithm~\cite{deb2002fast}, utilizing measured terminal voltages and surface temperatures as reference data, to determine the parameters of a full-order DFN model incorporating thermal effects. Rahman et al.~\cite{rahman2016electrochemical} utilized the PSO to estimate the electrode diffusion coefficients and reaction rate constants for a ROM. These population-based identification approaches are readily implementable due to their dependence on forward predictions alone. However, their stochastic search mechanisms demand numerous evaluations of forward problems, resulting in substantial computational costs. This challenge becomes more pronounced in high-dimensional parameter spaces due to the exponential growth of the search space. In contrast, gradient-based identification methods leverage the sensitivity information (gradient of the objective function with respect to parameters) to determine the search direction, thereby substantially enhancing the convergence rate and accelerating the optimization process.

However, due to the high nonlinearity and multi-scale nature of the DFN model, accurately computing sensitivity information poses significant challenges. The methods for computing sensitivity information in relevant studies include finite difference methods (FDM)~\cite{deng2017implementation,jin2018parameter}, complex-step methods~\cite{xue2013optimization, du2013optimization}, and analytical methods~\cite{masoudi2015parameter,scharrer2013new}. Deng et al.~\cite{deng2017implementation} employed the FDM to compute the sensitivity of the output voltage with respect to model parameters and utilized the Levenberg-Marquardt (L-M) method~\cite{more2006levenberg} to achieve parameter identification for a ROM. Similarly, Jin et al.~\cite{jin2018parameter} adopted the FDM and the L-M method to identify the thermodynamic and kinetic parameters of the full-order DFN model. In addition, the complex-step method, similar to the FDM, was employed by Xue et al.~\cite{xue2013optimization} and Du et al.~\cite{du2013optimization} to compute the parameter sensitivity of the full-order DFN model, with applications in the design optimization of battery cells. The advantage of the FDM and the complex-step method lies in their implementation simplicity, without the need for additional derivations. Nevertheless, these approaches often entail substantial computational costs, especially for inverse problems involving large parameter spaces. A more efficient alternative is to determine the sensitivity through analytical derivations. Masoudi et al.~\cite{masoudi2015parameter} computed parameter sensitivities by solving the sensitivity equations obtained through differentiation of the governing equations and employed a homotopy optimization method to identify parameters in the full-order DFN model. Scharrer et al.~\cite{scharrer2013new} proposed a space mapping approach to accelerate parameter identification, where the parameter sensitivity was derived using the adjoint method~\cite{cao2003adjoint,alberdi2018unified,givoli2021tutorial}. Validation results demonstrated that the adjoint method reduced computational cost by $20\%$ compared to the FDM. While the analytical methods enable more efficient and precise sensitivity computations, their implementation depends on intricate mathematical derivations and problem-specific adaptations. Consequently, developing efficient and convenient sensitivity computation methods remains to be a critical challenge in the parameter identification of LIBs.

Emerging differentiable programming techniques have demonstrated the capability to perform end-to-end gradient computation of various complicated physical systems~\cite{hu2019difftaichi, schoenholz2020jax, xue2023jax, mozaffar2023differentiable}, which offers a promising paradigm for developing advanced sensitivity computation methods for LIBs. 
In this work, we propose a differentiable LIB simulation framework called \texttt{DiffLiB} to bridge the gap between efficient parameter identification and computational challenge of sensitivity information. 
In \texttt{DiffLiB}, we have developed customized automatic differentiation (AD)~\cite{griewank2008evaluating} methods that can efficiently handle time-dependent DFN models involving strong nonlinearity with a multi-scale nature.
The key innovation is to identify the so-called VJP (vector-Jacobian product) structure in the chain of differentiation and define customized rules through the adjoint-based implicit differentiation methods~\cite{blondel2022efficient}.
\texttt{DiffLiB} enables end-to-end differentiability that not only facilitates parameter identification but is also potentially applicable to other inverse problems, such as battery design optimization~\cite{roy2022topology,guibert2024thermo}. 
Furthermore, the implementation of the differentiable full-order DFN model is under the finite element method (FEM)~\cite{hughes2003finite} that enables 3D high-fidelity modeling with real battery geometries. 
This feature establishes the foundation for quantifying the spatial inhomogeneity of state variables within LIBs and performing parameter identification with more realistic battery models.

The remainder of this paper is organized as follows: Section~\ref{sec:problem} introduces the formulations of the full-order DFN model, including the governing equations as well as the initial and boundary conditions. Section~\ref{Sec:fem} illustrates the weak formulations and the discretization scheme in time and space. Section~\ref{Sec:difflib} introduces the details of the proposed differentiable LIB simulation framework \texttt{DiffLiB}. 
Numerical examples are given in Section~\ref{Sec:examples} to demonstrate the accuracy and efficiency of \texttt{DiffLiB} in solving both forward prediction and inverse parameter identification problems. Finally, we conclude this work with possible extensions in Section~\ref{Sec:conclusion}. 
The source code of \texttt{DiffLiB} is available at \href{https://github.com/xwpken/DiffLiB}{https://github.com/xwpken/DiffLiB}.

\section{Problem formulations} 
\label{sec:problem}
\begin{figure}[H]
    \centering
    \includegraphics[width=1\linewidth]{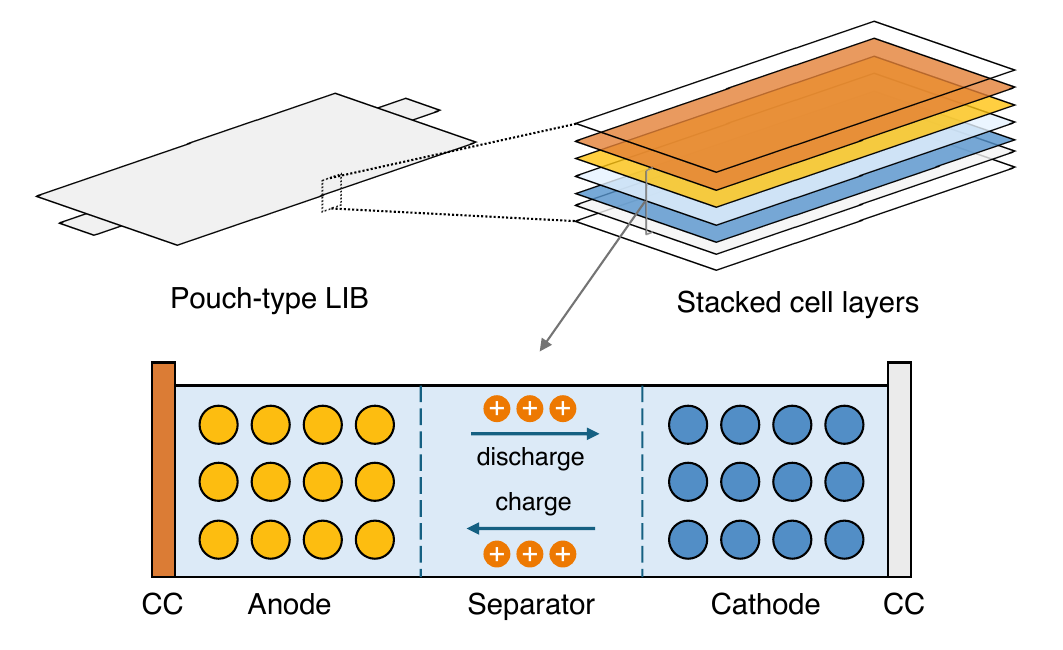}
    \caption{Schematic of the pouch-type lithium-ion battery.}
    \label{fig:LIB}
\end{figure}

As illustrated in Fig.~\ref{fig:LIB}, a pouch-type LIB consists of multiple stacked cell layers. Each layer comprises two electrodes (anode and cathode) and a porous separator, all of which immersed in a liquid electrolyte. The electrodes are composed of porous matrices with active particles: the anode typically uses graphite or silicon-based materials, while the cathode employs lithium compounds such as $\rm LiCoO_2$ or $\rm LiFePO_4$. Metal current collectors (denoted as CC), generally copper for the anode and aluminum for the cathode, are positioned at the outer ends of the electrodes to facilitate efficient charge transport.

According to the porous electrode theory, the DFN model simplifies the electrode as a composite of equal-size spherical particles and liquid electrolyte. The electrochemical reactions within the LIB cell are described through a system of governing equations across multiple scales: the microscopic scale within the active particles and the macroscopic scale of the cell layer. In most applications, the implemented DFN model adopts a 1D macroscopic geometry along the cell thickness, together with the microscopic dimension to form the so-called pseudo-two-dimensional (P2D) model. In this section, we extend the governing equations to more general P3D or P4D formulations, enabling the modeling with real battery geometries.

\subsection{Governing equations}
\label{sec:eqs}
For clarity, we first define the notation for different regions in the LIB cell. For the $n$-dimensional cell domain $\Omega\subset\mathbb{R}^{n}$, the subdomain with the electrolyte is defined as:
\begin{align}\label{domain_e}
    \Omega_{\textrm{e}} = \Omega_{\textrm{a}}\cup\Omega_{\textrm{se}}\cup\Omega_{\textrm{c}}
\end{align}
where $\Omega_{\textrm{a}}$, $\Omega_{\textrm{se}}$, and $\Omega_{\textrm{c}}$ are the subdomains of anode, separator, and cathode, respectively. The same subscript will be used to distinguish physical quantities from different regions. The solid phase include two parts: the active particles inside the electrode and the metal current collector attached to each electrode, which is denoted as:
\begin{align}
    \Omega_{\textrm{s}} = \Omega_{\textrm{s,a}}\cup\Omega_{\textrm{s,c}}
\end{align}
with
\begin{align}
\Omega_{\textrm{s},i} = 
\begin{cases}
        \Omega_{i},& n=2\\
         \Omega_{\textrm{cc},i}\cup\Omega_{i}, & n=3 
\end{cases}
\quad i\in\{\textrm{a},\textrm{c}\}
\end{align}
where $\Omega_{\textrm{cc},i}$ denote the subdomain of the electrode current collector. The microscopic domain of active particles is represented as $\Omega_{\textrm{r}} = \left[0,r_{\textrm{s}}\right]$, where $r_{\textrm{s}}$ is the particle radius. The time domain is set as $t\in\left[0,t_{\textrm{end}}\right]$.

During discharge, lithium ions diffuse from the interior to the surface of anode particles and subsequently deintercalate into the electrolyte. These lithium ions then migrate through the electrolyte to the cathode, where they combine with electrons transported through the external circuit to re-intercalate into cathode particles. This process reverses during charging, with lithium migrating from the cathode back to the anode. The above microscopic lithium diffusion is described by the Fick's second law in spherical coordinates:
\begin{align} \label{eq_cs}
    \frac{\partial c_{\textrm{s}}}{\partial t}-\frac{1}{r^2}\frac{\partial}{\partial r}\left(r^2D_{\textrm{s}}\frac{\partial c_{\textrm{s}}}{\partial r}\right)=0\quad &\textrm{in} \, \, \Omega_{\textrm{r}}\times\left[0,t_{\textrm{end}}\right]
\end{align}
where $c_{\textrm{s}}$ is the lithium concentration within the particle, $r$ is the microscopic spatial variable, and $D_{\textrm{s}}$ is the microscopic diffusivity. The lithium migration in the electrolyte is described by:
\begin{align}
\label{eq_ce}
    \varepsilon_{\textrm{e}}\frac{\partial c_{\textrm{e}}}{\partial t}-\nabla\cdot\left(D^{\textrm{eff}}_{\textrm{e}}\nabla c_{\textrm{e}}\right)=a_{\textrm{s}}\left(1-t_{\textrm{+}}\right)j_{\textrm{s}}\quad &\textrm{in} \, \, \Omega_{\textrm{e}}\times\left[0,t_{\textrm{end}}\right]
\end{align}
where $c_{\textrm{e}}$ is the electrolyte lithium concentration, $\varepsilon_{\textrm{e}}$ is the electrolyte volume fraction (porosity), and $D_{\textrm{e}}^{\textrm{eff}}$ is the effective electrolyte diffusivity. $a_{\textrm{s}}={3\varepsilon_{\textrm{s}}}/{r_{\textrm{s}}}$ represents the specific surface area of active particle, where $\varepsilon_{\textrm{s}}$ is the active particle volume fraction. $t_{\textrm{+}}$ is the transference number. $j_{\textrm{s}}$ is the pore wall flux to quantify the rate of lithium deintercalation and intercalation at the electrolyte-particle interface, which is governed by the Butler-Volmer (BV) equation \cite{newman2021electrochemical, dickinson2020butler}:
\begin{align}\label{eq_bv}
j_{\textrm{s}} = \frac{2i_{0}}{F}\sinh\left(\frac{F}{2RT}\eta\right)\quad \textrm{in} \, \, \Omega_{i}\times\left[0,t_{\textrm{end}}\right],\quad i\in\{\textrm{a},\textrm{c}\}
\end{align}
where $F$ is the Faraday constant, $R$ is the gas constant, and $T$ is the temperature. The exchange current density $i_{0}$ is computed as:
\begin{align}\label{eq_bv_i0}
i_{0} = k_{\textrm{s}} F\left(c_{\textrm{s}}^{\textrm{surf}}\right)^{1/2}\left(c_{\textrm{s}}^{\textrm{max}}-c_{\textrm{s}}^{\textrm{surf}}\right)^{1/2}\left(c_{\textrm{e}}\right)^{1/2}
\end{align}
where $k_{\textrm{s}}$ is the reaction rate constant. $c_{\textrm{s}}^{\textrm{surf}}$ is the microscopic lithium concentration at the particle surface. $c_{\textrm{s}}^{\textrm{max}}$ is the maximum intercalated lithium concentration of the electrode. The over-potential $\eta$ is defined as:
\begin{align}\label{eq_eta}
\eta = \phi_{\textrm{s}} - \phi_{\textrm{e}} - U_{\textrm{oc}}
\end{align}
where $\phi_{\textrm{s}}$ is the solid-phase potential. $\phi_{\textrm{e}}$ is the electrolyte potential. $U_{\textrm{oc}}$ is the open circuit potential (OCP) modeled as a function of $\frac{c_{\textrm{s}}^{\textrm{surf}}}{c_{\textrm{s}}^{\textrm{max}}}$.

With the deintercalation, migration, and intercalation of lithium, the resulting ionic and electronic fluxes generate the electric current. This transport process is described by the charge conservation equations for both the electrolyte and the solid phase, where the electrolyte potential $\phi_e$ is governed by:
\begin{align}\label{eq_phil}
\nabla\cdot\left[-\kappa^{\textrm{eff}}\nabla\phi_{\textrm{e}} + \frac{2RT\left(1-t_{\textrm{+}}\right)}{F}\kappa^{\textrm{eff}}\nabla\ln c_{\textrm{e}}\right] = a_{\textrm{s}}Fj_{\textrm{s}}\quad &\textrm{in} \, \, \Omega_{\textrm{e}}\times\left[0,t_{\textrm{end}}\right]
\end{align}
where $\kappa^{\textrm{eff}}$ is the effective electrolyte conductivity. The right-hand sides of (\ref{eq_ce}) and (\ref{eq_phil}) are eliminated in $\Omega_{\textrm{se}}$ because there are no active particles present. The charge conservation in the solid phase is expressed as:
\begin{align}\label{eq_phis}
\nabla\cdot\left(\sigma^{\textrm{eff}}\nabla\phi_{\textrm{s}}\right) = a_{\textrm{s}}Fj_{\textrm{s}}\quad &\textrm{in} \, \, \Omega_{\textrm{s}}\times\left[0,t_{\textrm{end}}\right]
\end{align}
where $\sigma^{\textrm{eff}}$ is the effective solid-phase conductivity. The above effective transfer coefficients $D_{\textrm{e}}^{\textrm{eff}}$, $\kappa^{\textrm{eff}}$, and $\sigma^{\textrm{eff}}$ are obtained from the Bruggeman relationship \cite{newman2021electrochemical, tjaden2016origin}:
\begin{align}
\label{eq_de_eff}
D_{\textrm{e}}^{\textrm{eff}} = \varepsilon_{\textrm{e}}^{\beta}D_{\textrm{e}},\,\, \kappa^{\textrm{eff}} = \varepsilon_{\textrm{e}}^{\beta}\kappa,\,\, \sigma^{\textrm{eff}} = \varepsilon_{\textrm{s}}^{\beta}\sigma
\end{align}
where $\beta$ is the Bruggeman coefficient. $D_{\textrm{e}}$ and $\kappa$ are the bulk diffusivity and conductivity of the electrolyte, respectively. $\sigma$ is the bulk conductivity of the solid phase . For the current collector, $\sigma^{\textrm{eff}}$ equals the bulk conductivity of the employed metal material, and the right-hand side of (\ref{eq_phis}) is also eliminated. 

\subsection{Initial and boundary conditions}
\label{sec:bcs}
For the time-dependent lithium diffusion (\ref{eq_cs}) and (\ref{eq_ce}), the initial conditions for $c_\textrm{s}$ and $c_{\textrm{e}}$ are set as:
\begin{align}\label{init_cs}
    c_{\textrm{s}}=c_{\textrm{s}}^0\quad &\textrm{on}\, \, \Omega_{\textrm{r}}\times\left\{0\right\} \\
    c_{\textrm{e}}=c_{\textrm{e}}^0\quad &\textrm{on}\, \, \Omega_{\textrm{e}}\times\left\{0\right\} 
\end{align}
where $c_{\textrm{s}}^0$ and $c_{\textrm{e}}^0$ are the initial value of $c_{\textrm{s}}$ and $c_{\textrm{e}}$, respectively. 

The boundary conditions for the microscopic lithium diffusion (\ref{eq_cs}) are set as:
\begin{align}\label{bcs_cs}
    D_{\textrm{s}}\frac{\partial c_{\textrm{s}}}{\partial r}=0\quad &\textrm{on}\, \, (r=0)\times\left[0,t_{\textrm{end}}\right] \nonumber \\
    D_{\textrm{s}}\frac{\partial c_{\text{s}}}{\partial r}=-j_{\textrm{s}}\quad &\textrm{on}\, \, (r=r_{\textrm{s}})\times\left[0,t_{\textrm{end}}\right] 
\end{align} 
where the surface Neumann boundary condition dynamically couples the microscopic diffusion (\ref{eq_cs}) with the interfacial reaction (\ref{eq_bv}), further connecting to the macroscopic reactions. 

The electrolyte subdomain is subject to zero-flux boundary conditions for both $c_{\textrm{e}}$ and $\phi_{\textrm{e}}$, which can be expressed as:
\begin{align}\label{bcs_ce_p}
    \nabla c_{\textrm{e}}\cdot\boldsymbol{n}=0\quad &\textrm{on} \, \, \partial\Omega_{\textrm{e}}\times\left[0,t_{\textrm{end}}\right] \\
    \nabla \phi_{\textrm{e}}\cdot\boldsymbol{n}=0\quad &\textrm{on} \, \, \partial\Omega_{\textrm{e}}\times\left[0,t_{\textrm{end}}\right] 
\end{align} 
where $\partial\Omega_{\textrm{e}}$ is the electrolyte subdomain boundary with the outward normal vector $\boldsymbol{n}$. 

For the charge conservation equation (\ref{eq_phil}) and (\ref{eq_phis}), the anode current collector tab $\partial\Omega_{\textrm{s,a}}^{\textrm{tab}}$ is grounded to provide a reference potential:
\begin{align}
    \phi_{\textrm{s}} = 0\quad \textrm{on}\, \,\partial\Omega_{\textrm{s,a}}^{\textrm{tab}}\times\left[0,t_{\textrm{end}}\right] 
\end{align}
while the external current with density $i_{\textrm{app}}$ is applied on the cathode current collector tab $\Omega_{\textrm{s,c}}^{\textrm{tab}}$:
\begin{align}
    \nabla\left(\sigma^{\textrm{eff}}\phi_{\textrm{s}}\right)\cdot\boldsymbol{n} = i_{\textrm{app}}\quad \textrm{on}\, \,\partial\Omega_{\textrm{s,c}}^{\textrm{tab}}\times\left[0,t_{\textrm{end}}\right] 
\end{align}
The remaining boundaries of $\Omega_{\textrm{s}}$ are treated as electrically insulating:
\begin{align}
    \nabla\phi_{\textrm{s}}\cdot\boldsymbol{n} = 0\quad \textrm{on}\, \,\partial\Omega_{\textrm{s}}\setminus(\partial\Omega_{\textrm{s,a}}^{\textrm{tab}}\cup\partial\Omega_{\textrm{s,c}}^{\textrm{tab}})\times\left[0,t_{\textrm{end}}\right] 
\end{align}
The terminal voltage can be computed as the difference of $\phi_{\textrm{s}}$ between the cathode and anode tabs:
\begin{align}
    V = \phi_{\textrm{s,c}}^{\textrm{tab}} - \phi_{\textrm{s,a}}^{\textrm{tab}}
\end{align}
If $\Omega\subset\mathbb{R}^2$, $\partial\Omega_{\textrm{s},i}^{\textrm{tab}}$ will be replaced by corresponding boundary parts of $\Omega_{i}$. 

\section{Weak form and discretization}
\label{Sec:fem}
The governing equations in subsection~\ref{sec:eqs}, together with the initial and boundary conditions in subsection~\ref{sec:bcs}, constitute the partial differential-algebraic equation (PDAE) system of the DFN model, requiring further reformulation and discretization to obtain the numerical solution. Two principle discretization techniques are the method of lines (MOL)~\cite{schiesser2012numerical} and the Rothe method~\cite{rothe1930zweidimensionale}. The MOL discretizes only the spatial variable while retaining the time variable as continuous, leading to differential-algebraic equations (DAEs). Specialized solvers such as SUNDIALS~\cite{hindmarsh2005sundials} and PETSc/TS~\cite{abhyankar2018petsc} can be applied to solve the obtained DAE system. The Rothe method first discretizes the time variable and then the spatial variable, resulting in algebraic equations that can be solved using general nonlinear solvers such as the Newton's method~\cite{kelley2003solving}.

In this study, we employ the Rothe method with the backward Euler discretization in time and the finite element discretization in space. We first define the trial and test function spaces as follows:
\begin{align}
    &\mathscr{S}_{\textrm{r}} = \left\{c_{\textrm{s}}\in H^1(\Omega_{\textrm{r}})\right\},\,\, \mathscr{T}_{\textrm{r}} = \left\{\hat{c}_{\textrm{s}}\in H^1(\Omega_{\textrm{r}})\right\}, \nonumber\\
    &\mathscr{S}_{\textrm{c}} = \left\{c_{\textrm{e}}\in H^1(\Omega_{\textrm{e}})\right\},\,\, \mathscr{T}_{\textrm{c}} = \left\{\hat{c}_{\textrm{e}}\in H^1(\Omega_{\textrm{e}})\right\},\nonumber\\
    &\mathscr{S}_{\textrm{p}} = \left\{\phi_{\textrm{e}}\in H^1(\Omega_{\textrm{e}})\right\},\,\, \mathscr{T}_{\textrm{p}} = \left\{\hat{\phi}_{\textrm{e}}\in H^1(\Omega_{\textrm{e}})\right\},\nonumber\\
    &\mathscr{S}_{\textrm{s}} = \left\{\phi_{\textrm{s}}\in H^1(\Omega_{\textrm{s}})\right\},\,\, \mathscr{T}_{\textrm{s}} = \left\{\hat{\phi}_{\textrm{s}}\in H^1(\Omega_{\textrm{s}})\mid\hat{\phi}_{\textrm{s}}=0\,\,\textrm{on}\,\,\partial\Omega_{\textrm{s,a}}^{\textrm{tab}}\right\}
\end{align}
where $\hat{\left(\cdot\right)}$ represents the test function. In the Sobolev space $H^1$, both the functions and their first-order weak derivatives are square integrable over the domain of definition. Then the weak form of the partial differential equations (PDEs) within the PDAE system can be formulated as:
\begin{itemize}
    \item Microscopic lithium concentration $c_{\textrm{s}}$:
    Find $c_{\textrm{s}}\in \mathscr{S}_{\textrm{r}}$, such that
    \begin{align}
        \int_{0}^{r_{\textrm{s}}}r^2\hat{c}_{\textrm{s}}\frac{\partial c_{\textrm{s}}}{\partial t}\textrm{d}r+\int_{0}^{r_{\textrm{s}}}r^2D_{\textrm{s}}\frac{\partial\hat{c}_{\textrm{s}}}{\partial r}\frac{\partial c_{\textrm{s}}}{\partial r}\textrm{d}r+r_{\textrm{s}}^2\hat{c}_{\textrm{s}}\left(r_{\textrm{s}}\right)j_{\textrm{s}}=0,\quad \forall\hat{c}_{\textrm{s}}\in \mathscr{T}_{\textrm{r}}
    \end{align}

    \item Electrolyte lithium concentration $c_{\textrm{e}}$:
    Find $c_{\textrm{e}}\in \mathscr{S}_{\textrm{c}}$, such that
    \begin{align}
        \int_{\Omega_{\textrm{e}}}\varepsilon_{\textrm{e}}\hat{c}_{\textrm{e}}\frac{\partial c_{\textrm{e}}}{\partial t}\textrm{d}V+
        \int_{\Omega_{\textrm{e}}}D_{\textrm{e}}^{\textrm{eff}}\nabla\hat{c}_{\textrm{e}}\cdot\nabla c_{\textrm{e}}\textrm{d}V-\int_{\Omega_{\textrm{e}}}a_{\textrm{s}}(1-t_{\textrm{+}})\hat{c}_{\textrm{e}}j_{\textrm{s}}\textrm{d}V=0,\quad \forall\hat{c}_{\textrm{e}}\in \mathscr{T}_{\textrm{c}}
    \end{align}

    \item Electrolyte potential $\phi_{\textrm{e}}$:
    Find $\phi_{\textrm{e}}\in \mathscr{S}_{\textrm{p}}$, such that
    \begin{align} \int_{\Omega_{\textrm{e}}}\kappa^{\textrm{eff}}\nabla\hat{\phi}_{\textrm{e}}\cdot\nabla\phi_{\textrm{e}}\textrm{d}V-\int_{\Omega_{\textrm{e}}}\frac{2RT\left(1-t_{+}\right)}{F}\kappa^{\textrm{eff}}\nabla\hat{\phi}_{\textrm{e}}\cdot\nabla\ln c_{\textrm{e}}\textrm{d}V-\int_{\Omega_{\textrm{e}}}a_{\textrm{s}}F\hat{\phi}_{\textrm{e}}j_{\textrm{s}}\textrm{d}V=0,\quad \forall\hat{\phi}_{\textrm{e}}\in \mathscr{T}_{\textrm{p}}
    \end{align}

    \item Solid-phase potential $\phi_{\textrm{s}}$:
    Find $\phi_{\textrm{s}}\in \mathscr{S}_{\textrm{s}}$, such that
    \begin{align}
        \int_{\Omega_{\textrm{s}}}\sigma^{\textrm{eff}}\nabla\hat{\phi}_{\textrm{s}}\cdot\nabla\phi_{\textrm{s}}\textrm{d}V+\int_{\Omega_{\textrm{s}}}a_{\textrm{s}}F\hat{\phi}_{\textrm{s}}j_{\textrm{s}}\textrm{d}V-\int_{\partial\Omega_{\textrm{s,c}}^{\textrm{tab}}}\hat{\phi}_{\textrm{s}}i_{\textrm{app}}\textrm{d}A=0,\quad \forall\hat{\phi}_{\textrm{s}}\in \mathscr{T}_{\textrm{s}}
    \end{align}
    
\end{itemize}

The weak form is first discretized in time via an implicit backward Euler scheme, where the temporal derivative of a state variable $w$ is approximated as:
\begin{align}\label{dis_time}
    \frac{\partial w}{\partial t} = \frac{w^{i}-w^{i-1}}{\Delta t}
\end{align}
where $\Delta t$ is the time step size. With the finite element discretization in space, $w$ can be expressed as the interpolation of the nodal value $w_k$ with the shape function $\psi_k$:
\begin{align}\label{dis_fem}
    w = \sum_{k=1}^{K}\psi_kw_k
\end{align}
Here, we employ the standard Galerkin approximation, and the test function $\hat{w}$ adopts the same interpolation scheme as (\ref{dis_fem}). Note that the microscopic computation is governed by the macroscopic variable $j_s$, we define the microscopic mesh at each macroscopic mesh node within the electrode subdomain. For the $k$-th macroscopic mesh node, the microscopic residual vector is defined as:
\begin{align}\label{res_mic}
    \boldsymbol{R}^i_{\textrm{mic},k} = [R^i_{\textrm{r},1},\,\cdots,\,R^i_{\textrm{r},n},\cdots ,R^i_{\textrm{r},M_{\textrm{r}}}]^{\textrm{T}}
\end{align}
with the $n$-th component as:
\begin{align}\label{res_cs}
    R^i_{\textrm{r},n} = \int_{0}^{r_{\textrm{s}}}r^2\psi_n\frac{c_{\textrm{s},m}^{i}-c_{\textrm{s},m}^{i-1}}{\Delta t}\psi_m\textrm{d}r+\int_{0}^{r_{\textrm{s}}}r^2D_{\textrm{s}}\frac{\partial \psi_{n}}{\partial r}\frac{\partial \psi_{m}}{\partial r}c^i_{\textrm{s},m}\textrm{d}r+r_{\textrm{s}}^2\psi_{n}\left(r_{\textrm{s}}\right)j^i_{\textrm{s},k}
\end{align}
where $M_{\textrm{r}}$ is the number of microscopic mesh nodes. The composition of the macroscopic residual vector is determined by the variables defined in the current domain. Here, we consider the electrode subdomain encompassing all the macroscopic variables as an example, where the residual vector of the $k$-th node can be expressed as:
\begin{align}\label{res_mac}
    \boldsymbol{R}^i_{\textrm{mac},k} = [R^i_{\textrm{c},k},\,R^i_{\textrm{p},k},\,R^i_{\textrm{s},k},\,R^i_{\textrm{j},k}]^{\textrm{T}}
\end{align}
where
\begin{align} \label{res_c}
    R^i_{\textrm{c},k} = \int_{\Omega_{\textrm{e}}}\varepsilon_{\textrm{e}}\psi_k\frac{c_{\textrm{e},m}^{i}-c_{\textrm{e},m}^{i-1}}{\Delta t}\psi_m\textrm{d}V+\int_{\Omega_{\textrm{e}}}D_{\textrm{e}}^{\textrm{eff}}\frac{\partial \psi_{k}}{\partial x_q}\frac{\partial \psi_{m}}{\partial x_q}c^i_{\textrm{e},m}\textrm{d}V-\int_{\Omega_{\textrm{e}}}a_{\textrm{s}}(1-t_{\textrm{+}})\psi_{k}\psi_{m}j^i_{\textrm{s},m}\textrm{d}V
\end{align}

\begin{align} \label{res_p}
    R^i_{\textrm{p},k} = \int_{\Omega_{\textrm{e}}}\kappa^{\textrm{eff}}\frac{\partial \psi_{k}}{\partial x_q}\frac{\partial \psi_{m}}{\partial x_q}\phi^i_{\textrm{e},m}\textrm{d}V-\int_{\Omega_{\textrm{e}}}\frac{2RT(1-t_{\textrm{+}})}{F}\frac{\kappa^{\textrm{eff}}}{\psi_{n}c^i_{\textrm{e},n}}\frac{\partial \psi_{k}}{\partial x_q}\frac{\partial \psi_{m}}{\partial x_q}c^i_{\textrm{e},m}\textrm{d}V-\int_{\Omega_{\textrm{e}}}a_{\textrm{s}}F\psi_{k}\psi_{m}j^i_{\textrm{s},m}\textrm{d}V
\end{align}

\begin{align} \label{res_s}
    R^i_{\textrm{s},k} = \int_{\Omega_{\textrm{s}}}\sigma^{\textrm{eff}}\frac{\partial \psi_{k}}{\partial x_q}\frac{\partial \psi_{m}}{\partial x_q}\phi^i_{\textrm{s},m}\textrm{d}V+\int_{\Omega_{\textrm{s}}}a_{\textrm{s}}F\psi_{k}\psi_mj^i_{\textrm{s},m}\textrm{d}V - \int_{\partial\Omega_{\textrm{s,c}}^{\textrm{tab}}}\psi_ki_{\textrm{app}}\textrm{d}A
\end{align}

\begin{align} \label{res_j}
        R^i_{\textrm{j},k} = j^i_{\textrm{s},k} - \frac{2i_0(c_{\textrm{s},k}^{\textrm{surf},i},\,c^i_{\textrm{e},k})}{F}\sinh\left[\frac{F}{2RT}\eta\left(c_{\textrm{s},k}^{\textrm{surf},i},\,\phi^i_{\textrm{e},k},\,\phi^i_{\textrm{s},k}\right)\right]
\end{align}

\section{Differentiable LIB simulation framework}
\label{Sec:difflib}
This section gives a detailed explanation of the proposed differentiable LIB simulation framework \texttt{DiffLiB}, including the solution of forward and inverse problems. With the backend support from the modern scientific computing framework \texttt{JAX} \cite{jax2018github} and the differentiable FEM library \texttt{JAX-FEM} \cite{xue2023jax}, \texttt{DiffLiB} enables the automatic computation of the Jacobian matrix in forward problems. A sequential solution scheme is introduced to reduce the size of the linear equation system in the multi-scale computations. For the inverse problem, customized differentiation rules are defined to enable the automatic computation of time-dependent sensitivity information via AD.

\subsection{The forward problem}
\label{Sec:fwd}
Let $N$ be the number of discrete time steps after temporal discretization, and $M$ the total number of microscopic and macroscopic degrees of freedom (DOF) after spatial discretization. The residual vectors (\ref{res_mic}) to (\ref{res_j}) form the following nonlinear system at the $i$-th time step:
\begin{align}\label{eq_Ri}
       \boldsymbol{R}^i\left(\boldsymbol{U}^i,\boldsymbol{U}^{i-1},\boldsymbol{\theta}\right) = \boldsymbol{0},\quad i=1,\cdots,N
\end{align}
where $\boldsymbol{U}^i\in\mathbb{R}^M$ is the solution vector to be solved: 
\begin{align}\label{eq_U}
\boldsymbol{U}^i = \left[\boldsymbol{U}_{\textrm{r}}^i,\boldsymbol{U}_{\textrm{c}}^i,\boldsymbol{U}_{\textrm{p}}^i,\boldsymbol{U}_{\textrm{s}}^i,\boldsymbol{U}_{\textrm{j}}^i\right]^{\textrm{T}}
\end{align}
with the components representing the vector of nodal values of discretized $c_{\textrm{s}}$, $c_{\textrm{e}}$, $\phi_{\textrm{e}}$, $\phi_{\textrm{s}}$, and $j_{\textrm{s}}$, respectively. $\boldsymbol{U}^{i-1}$ is the previous solution vector. The input parameter vector $\boldsymbol{\theta}\in\mathbb{R}^L$ incorporates the original DFN model parameters (e.g., $c_{\textrm{s}}^{\textrm{max}}$ and $c_{\textrm{s}}^{0}$) as well as other customized parameters. $\boldsymbol{R}^i:\mathbb{R}^M\times\mathbb{R}^M\times\mathbb{R}^L\rightarrow\mathbb{R}^M$ is the residual function dependent on $\boldsymbol{U}^i$, $\boldsymbol{U}^{i-1}$, and $\boldsymbol{\theta}$. In \texttt{DiffLiB}, we adopt the Newton's method to linearize (\ref{eq_Ri}):
\begin{align}\label{eq_NR}
       \boldsymbol{\mathcal{J}}\vert_{n}^{i}\Delta\boldsymbol{U}\vert_{n+1}^{i} = -\boldsymbol{R}\vert_n^i
\end{align}
where $\left(\cdot\right)\vert_n^i$ denotes the quantity of the $n$-th Newton iteration in the computation of the $i$-th time step. $\boldsymbol{\mathcal{J}}\vert_{n}^{i}\in\mathbb{R}^{M\times M}$ is the Jacobian matrix evaluated at $\boldsymbol{U}\vert_{n}^{i}$ through the tangent function $\frac{\partial\boldsymbol{R}}{\partial\boldsymbol{U}}$. The incremental solution vector $\Delta\boldsymbol{U}\vert_{n+1}^{i}$ is added on $\boldsymbol{U}\vert_n^i$ to obtain $\boldsymbol{U}\vert_{n+1}^i$, followed by the next Newton iteration. This process is repeated until the residual vector satisfies the convergence criteria, and the final Newton iteration solution will be output as $\boldsymbol{U}^i$.

\subsubsection{Automatic computation of Jacobian}
\label{Sec:fwd:auto}
In traditional finite element implementations, the explicit expression for $\frac{\partial\boldsymbol{R}}{\partial\boldsymbol{U}}$ needs to be derived analytically, which is often an error-prone process. 
This inconvenience is especially pronounced in the DFN model and its extensions, which encompass numerous parameters and the coupling of multiple physical fields. In \texttt{DiffLiB}, we develop a flexible physical modeling module to eliminate the need for these cumbersome derivations, which enables the automatic computation of Jacobian matrix via AD.

Although AD has been developed for several decades in the field of machine learning, its application in differentiable simulation remains in early stages. Therefore, we first present some basic concepts to facilitate the understanding of AD, particularly in the context of the modern AD framework \texttt{JAX}. Consider a function $\boldsymbol{f}:\mathbb{R}^{M}\rightarrow\mathbb{R}^{N}$ composed of sequential differentiable operations $\boldsymbol{g}_1,\,\boldsymbol{g}_2,\,\cdots,\,\boldsymbol{g}_{n-1},\,\boldsymbol{g}_{n}$:
\begin{align}\label{eq_func}
    \boldsymbol{y} = \boldsymbol{f}\left(\boldsymbol{x}\right) = \boldsymbol{g}_n\circ \boldsymbol{g}_{n-1}\circ \cdots\circ \boldsymbol{g}_2\circ \boldsymbol{g}_1\left(\boldsymbol{x}\right)
\end{align}
where $\boldsymbol{x}\in\mathbb{R}^{M}$ and $\boldsymbol{y}\in\mathbb{R}^{N}$ are the input and output vectors, respectively. The Jacobian matrix of $\boldsymbol{f}$ evaluated at $\boldsymbol{x}$, denoted as $\boldsymbol{\mathcal{J}}\in\mathbb{R}^{N\times M}$, can be obtained with chain rules:
\begin{align}\label{eq_chain}
    \boldsymbol{\mathcal{J}} = \frac{\partial \boldsymbol{g}_n}{\partial\boldsymbol{u}_{n-1}} \frac{\partial \boldsymbol{g}_{n-1}}{\partial\boldsymbol{u}_{n-2}}\cdot\cdots\cdot\frac{\partial \boldsymbol{g}_2}{\partial\boldsymbol{u}_{1}}\frac{\partial \boldsymbol{g}_1}{\partial\boldsymbol{x}}
\end{align}
where $\boldsymbol{u}_i=\boldsymbol{g}_i(\boldsymbol{u}_{i-1})$ are the intermediate variables with $\boldsymbol{u}_0=\boldsymbol{x}$ and $\boldsymbol{u}_n=\boldsymbol{y}$. To compute $\boldsymbol{\mathcal{J}}$, AD first converts (\ref{eq_func}) into a computational graph \cite{griewank2008evaluating} with variables represented as nodes and operations as connecting edges. Each operation $\boldsymbol{g}_i$ can be decomposed into a combination of elementary operations (e.g., addition, multiplication, exponentiation), which have known derivative expressions. However, instead of explicitly computing each term in (\ref{eq_chain}), AD uses a seed vector to propagate along the computational graph and accumulate the contribution of each node to form $\boldsymbol{\mathcal{J}}$. This process can be executed from the inputs to the outputs or vice versa, which are known as forward-mode AD and reverse-mode AD, respectively.

The forward-mode AD starts by computing the directional derivative of $\boldsymbol{g}_1$ at $\boldsymbol{x}$ on the direction of a seed vector $\boldsymbol{v}\in\mathbb{R}^{M}$, which will continue to propagate in an input-to-output order:
\begin{align}\label{eq_fwd_ad}
    \dot{\boldsymbol{u}}_{i} = \frac{\partial \boldsymbol{g}_i}{\partial\boldsymbol{u}_{i-1}}\dot{\boldsymbol{u}}_{i-1}
\end{align}
where $\dot{\boldsymbol{u}}_{i}$ denotes the directional vector with $ \dot{\boldsymbol{u}}_{0} = \boldsymbol{v}$. The propagation results give the following Jacobian-vector product (JVP):
\begin{align}\label{eq_jvp}
    \textrm{D}\boldsymbol{f}\left(\boldsymbol{x}\right)\left[\boldsymbol{v}\right] = \boldsymbol{\mathcal{J}}\boldsymbol{v}
\end{align}
where $\textrm{D} \boldsymbol{f}\left(\boldsymbol{x}\right):\mathbb{R}^{M}\rightarrow\mathbb{R}^{N}$ denotes the Jacobian linear map of $\boldsymbol{f}$ at $\boldsymbol{x}$. Eq.~(\ref{eq_jvp}) can also be regarded as the directional derivative of $\boldsymbol{f}$ at $\boldsymbol{x}$ on the direction of $\boldsymbol{v}$. If $\boldsymbol{v}$ is chosen as the basis vector $\boldsymbol{e}_i\in\mathbb{R}^{M}$, the forward propagation will give the $i$-th column of $\boldsymbol{\mathcal{J}}$. Thus, the complete Jacobian matrix can obtained with $M$ forward propagation.

The reverse-mode AD, also known as the backpropagation \cite{rumelhart1986learning}, first performs a complete forward computation of (\ref{eq_func}) and stores all the intermediate variables. Then a seed vector $\boldsymbol{v}\in\mathbb{R}^{N}$ will propagate from the output to the input:
\begin{align}\label{eq_rev_ad}
    \left(\bar{\boldsymbol{u}}_{i-1}\right)^{\textrm{T}} = \left(\bar{\boldsymbol{u}}_i\right)^{\textrm{T}}\frac{\partial \boldsymbol{g}_i}{\partial \boldsymbol{u}_{i-1}}
\end{align}
where $\bar{\boldsymbol{u}}_i$ represents the adjoint vector with $\bar{\boldsymbol{u}}_{n} = \boldsymbol{v}$. The accumulation equals to the transpose of the following vector-Jacobian product (VJP):
\begin{align}\label{eq_vjp}
    \textrm{D}f\left(\boldsymbol{x}\right)^{*}\left[\boldsymbol{v}\right] = \boldsymbol{\mathcal{J}}^{\textrm{T}}\boldsymbol{v}
\end{align}
where $\textrm{D} \boldsymbol{f}\left(\boldsymbol{x}\right)^{*}:\mathbb{R}^{N}\rightarrow\mathbb{R}^{M}$ is the adjoint linear map of $\textrm{D}\boldsymbol{f}\left(\boldsymbol{x}\right)$. Similarly, we can get the $i$-th row of $\boldsymbol{\mathcal{J}}$ by selecting the seed vector as the basis vector $\boldsymbol{e}_i\in\mathbb{R}^{N}$. Therefore, the complete Jacobian matrix can be constructed with one forward computation and $N$ reverse propagation.

The efficiency of the forward and reverse mode of AD relies on the dimension of $\boldsymbol{x}$ and $\boldsymbol{y}$. When $M\ll N$, the forward mode is more efficient. When $M\gg N$
, the reverse mode exhibits higher efficiency. When $M$ is close to $N$, the efficiency of these two modes is almost the same. If the memory usage is taken into account, the forward-mode AD is preferred, as it does not require storing intermediate variables. For interested readers, we recommend a comprehensive discussion on the computational complexity of AD \cite{blondel2024elements}. 

In \texttt{DiffLiB}, the AD is used to compute two types of Jacobians: the $\boldsymbol{\mathcal{J}}\vert_n^i$ in the Newton iteration of the forward problem and the sensitivity information for the inverse problem, where the latter will be discussed in subsection \ref{Sec:inv}. For the forward problem, $\boldsymbol{\mathcal{J}}\vert_n^i$ is usually a sparse matrix due to the compact support property of the shape function in (\ref{dis_fem}). Directly constructing $\boldsymbol{\mathcal{J}}\vert_n^i$ via AD would lead to unnecessary computational cost for numerous zero elements. Therefore, we adopt the AD at the element level. We assume that the number of DOF for each element is $M_{\textrm{c}}$. The element-level Jacobian matrix $\boldsymbol{\mathcal{J}}_{\textrm{e}}\in\mathbb{R}^{M_{\textrm{c}}\times M_{\textrm{c}}}$ can be computed as follows:
\begin{align}\label{eq_cell_J}
    \boldsymbol{\mathcal{J}}_\textrm{e}\vert_n^i = \frac{\partial\boldsymbol{R}_\textrm{e}}{\partial\boldsymbol{U}_\textrm{e}}\bigg|_n^i
\end{align}
where $\boldsymbol{U}_e\in\mathbb{R}^{M_{\textrm{c}}}$ is vector of nodal DOF for each element. $\boldsymbol{R}_e:\mathbb{R}^{M_{\textrm{c}}}\times\mathbb{R}^{M_{\textrm{c}}}\times\mathbb{R}^L\rightarrow\mathbb{R}^{M_{\textrm{c}}}$ is the element-level residual function. Since $\boldsymbol{U}_e$ has the same dimension as the output of $\boldsymbol{R}_e$, we adopt the forward-mode AD to compute $\boldsymbol{\mathcal{J}}_e$. The vectorized operations in \texttt{JAX} enable batch computation of (\ref{eq_cell_J}) without using explicit for-loops, which can significant improve the computational efficiency. After that, the element-level Jacobian matrices will be assembled to form the global Jacobian matrix $\boldsymbol{\mathcal{J}}\vert_n^i$ for the Newton iteration. The above process is fully automated and can be naturally extended to more sophisticated coupling mechanisms through modifications to the implemented weak formulations.

\subsubsection{Multi-scale solution scheme}
\label{Sec:sol}
Since a microscopic mesh is attached to each mesh node within the electrode subdomain, the total number of microscopic DOF is much greater than that of macroscopic DOF. Simultaneously solving for both macroscopic and microscopic solutions results in a large-scale linear system (\ref{eq_NR}), whose size grows significantly with mesh refinement. The weak formulations in Section~\ref{Sec:fem} reveals that among the macroscopic variables, only the pore wall flux $j_{\textrm{s}}$ affects the microscopic computation, while the macroscopic computation requires only the surface microscopic lithium concentration $c_{\textrm{s}}^{\textrm{surf}}$. Therefore, we introduce a sequential solution scheme to reduce the size of the linear system to be solved. 

For each Newton iteration, we first solve the microscopic problem at each mesh node within the electrode subdomain to obtain $c_{\textrm{s}}^{\textrm{surf}}$. We assume that all microscopic computational domains have the same mesh discretization with $M_{\textrm{r}}$ nodes, and the involved parameters (e.g., microscopic diffusivity $D_{\textrm{s}}$) remain constant during the computation. The microscopic linear system for the $k$-th macroscopic node is given by:
\begin{align}\label{eq_mic_inc}
\boldsymbol{\mathcal{J}}_{\textrm{mic},k}\left(\boldsymbol{U}_{\textrm{r},k}\vert_n^i-\boldsymbol{U}_{\textrm{r},k}^{i-1}\right)= -\boldsymbol{R}_{\textrm{mic},k}\vert_{n}^i
\end{align}
where the microscopic Jacobian matrix $\boldsymbol{\mathcal{J}}_{\textrm{mic},k}\in\mathbb{R}^{M_{\textrm{r}}\times M_{\textrm{r}}}$ is identical for all particles in the same electrode subdomain and remains constant throughout the computation. $\boldsymbol{U}_{\textrm{r},k}\in\mathbb{R}^{M_{\textrm{r}}}$ denotes the vector of microscopic DOF. With (\ref{res_cs}), the right-hand side $\boldsymbol{R}_{\textrm{mic},k}\in\mathbb{R}^{M_{\textrm{r}}}$ can be decomposed into two parts:
\begin{align}\label{eq_mic_rhs}
    \boldsymbol{R}_{\textrm{mic},k}\vert_{n}^i = \boldsymbol{r}_{\textrm{mic},k}^{\textrm{I}}\left(\boldsymbol{U}_{\textrm{r},k}^{i-1}\right)+\boldsymbol{r}_{\textrm{mic},k}^{\textrm{II}}\left(\boldsymbol{U}_{\textrm{j},k}\vert_{n}^i\right)
\end{align}
where the first part $\boldsymbol{r}_{\textrm{mic},k}^{\textrm{I}}\in\mathbb{R}^{M_{\textrm{r}}}$ depends on the previous microscopic solution $\boldsymbol{U}_{\textrm{r},k}^{i-1}\in\mathbb{R}^{M_{\textrm{r}}}$. The second part $\boldsymbol{r}_{\textrm{mic},k}^{\textrm{II}}\in\mathbb{R}^{M_{\textrm{r}}}$ relies on the pore wall flux $\boldsymbol{U}_{\textrm{j},k}\in\mathbb{R}$ at the $k$-th node, with non-zero value only at the index of particle surface position. Then, we can compute the surface lithium concentration for this newton iteration by:
\begin{align}\label{eq_mic_surf}
   c_{\textrm{s},k}^{\textrm{surf}}\vert_{n}^{i} = \boldsymbol{L}^{\textrm{T}}_{\textrm{surf}}\boldsymbol{U}_{\textrm{r},k}\vert_{n}^{i}
\end{align}
where $\boldsymbol{L}_{\textrm{surf}}\in\mathbb{R}^{M_{\textrm{r}}}$ is the locating vector with a value of $1$ at the index of particle surface position and 0 elsewhere.

Note that for the same electrode subdomain, only the right-hand side of (\ref{eq_mic_inc}) changes, while $\boldsymbol{\mathcal{J}}_{\textrm{mic},k}$ remains constant. In \texttt{DiffLiB}, we employ a similar approach as Han et al.~\cite{han2021fast} in their FDM-based study to simplify the microscopic computation, which can reduce the computational cost while maintain the same accuracy. We first solve the following adjoint equation to obtain the adjoint vector $\boldsymbol{\lambda}_{\textrm{surf},k}$:
\begin{align}\label{eq_adj_mic}
\boldsymbol{\mathcal{J}}_{\textrm{mic},k}^{\textrm{T}}\boldsymbol{\lambda}_{\textrm{surf},k} = \boldsymbol{L}_{\textrm{surf}}
\end{align}
which requires solving only once for each of the anode and cathode.
With (\ref{eq_mic_inc}) to (\ref{eq_mic_surf}), the surface lithium concentration can be computed by:
\begin{align}\label{eq_mic_surf_simp}
   c_{\textrm{s},k}^{\textrm{surf}}\vert_{n}^{i} = \boldsymbol{L}^{\textrm{T}}_{\textrm{surf}}\boldsymbol{U}_{\textrm{r},k}^{i-1}-\boldsymbol{\lambda}_{\textrm{surf},k}^{\textrm{T}}\boldsymbol{r}_{\textrm{mic},k}^{\textrm{I}}\left(\boldsymbol{U}_{\textrm{r},k}^{i-1}\right)-\boldsymbol{\lambda}_{\textrm{surf},k}^{\textrm{T}}\boldsymbol{r}_{\textrm{mic},k}^{\textrm{II}}\left(\boldsymbol{U}_{\textrm{j},k}\vert_{n}^i\right)
\end{align}
where the first two terms of the right-hand side can be precomputed at the beginning of each time step, and only the last term is updated during Newton iterations. Consequently, the microscopic computation is reduced to a few vector multiplications, which is much more efficient than directly solving (\ref{eq_mic_inc}) on each mesh node within the electrode subdomain.

After computing the microscopic problems, the macroscopic residual vector is evaluated and the corresponding Jacobian matrix is automatically computed via AD. We assume that the total number of macroscopic DOF is $M_{\textrm{k}}$. The macroscopic incremental solution can be obtained by solving:
\begin{align}\label{eq_mac_inc}
    \underbrace{\begin{bmatrix}
      \boldsymbol{\mathcal{J}}_{\textrm{cc}}&  \boldsymbol{\mathcal{J}}_{\textrm{cp}}&  \boldsymbol{\mathcal{J}}_{\textrm{cs}}& \boldsymbol{\mathcal{J}}_{\textrm{cj}}\\
        \boldsymbol{\mathcal{J}}_{\textrm{pc}}&  \boldsymbol{\mathcal{J}}_{\textrm{pp}}&  \boldsymbol{\mathcal{J}}_{\textrm{ps}}& \boldsymbol{\mathcal{J}}_{\textrm{pj}}\\
      \boldsymbol{\mathcal{J}}_{\textrm{sc}}&  \boldsymbol{\mathcal{J}}_{\textrm{sp}}&  \boldsymbol{\mathcal{J}}_{\textrm{ss}}& \boldsymbol{\mathcal{J}}_{\textrm{sj}}\\
      \boldsymbol{\mathcal{J}}_{\textrm{jc}}&  \boldsymbol{\mathcal{J}}_{\textrm{jp}}&  \boldsymbol{\mathcal{J}}_{\textrm{js}}& \boldsymbol{\mathcal{J}}_{\textrm{jj}}\\
    \end{bmatrix}_{n}^i}_{\boldsymbol{\mathcal{J}}_{\textrm{mac}}\vert_n^i}
    \underbrace{
    \begin{bmatrix}
      \Delta\boldsymbol{U}_{\textrm{c}}\\
      \Delta\boldsymbol{U}_{\textrm{p}}\\
      \Delta\boldsymbol{U}_{\textrm{s}}\\
      \Delta\boldsymbol{U}_{\textrm{j}}
    \end{bmatrix}_{n+1}^i}_{\Delta\boldsymbol{U}_{\textrm{mac}}\vert_{n+1}^i}
    =
    -\underbrace{\begin{bmatrix}
      \boldsymbol{R}_{\textrm{c}}\\
      \boldsymbol{R}_{\textrm{p}}\\
      \boldsymbol{R}_{\textrm{s}}\\
      \boldsymbol{R}_{\textrm{j}}
    \end{bmatrix}_{n}^i}_{\boldsymbol{R}_{\textrm{mac}}\vert_n^i}
\end{align}
where $\boldsymbol{\mathcal{J}}_{\textrm{mac}}\in\mathbb{R}^{M_{\textrm{k}}\times M_\textrm{k}}$ is the macroscopic Jacobian matrix, $\Delta\boldsymbol{U}_{\textrm{mac}}\in\mathbb{R}^{M_{\textrm{k}}}$ is the incremental solution vector, and $\boldsymbol{R}_{\textrm{mac}}\in\mathbb{R}^{M_{\textrm{k}}}$ is the residual vector evaluated by (\ref{res_c}) to (\ref{res_j}). With $\Delta\boldsymbol{U}_{\textrm{mac}}$, the updated $j_{\textrm{s}}$ is passed to the microscopic problem, which is then recomputed to yield a new $c_{\textrm{s}}^{\textrm{surf}}$. The above process is repeated until the convergence criteria of the Newton iteration are satisfied. We emphasize that, although the microscopic and macroscopic problems are solved sequentially, the computational accuracy remains equivalent to that of simultaneous solution, as both computations are still embedded within the same Newton iteration.

\subsection{The inverse problem}
\label{Sec:inv}
With the forward problem formulation in subsection \ref{Sec:fwd}, we model the inverse problem as the following discretized PDAE-constrained optimization problem:
\begin{align}\label{eq_inv}
    &\min_{\boldsymbol{\mathcal{U}}\in\mathbb{R}^{M\times N},\ \boldsymbol{\theta}\in\mathbb{R}^{L}}\mathcal{L}\left(\boldsymbol{\mathcal{U}}\left(\boldsymbol{\theta}\right),\boldsymbol{\theta}\right)\nonumber\\
    &\textrm{s.t.}\quad\boldsymbol{R}^i\left(\boldsymbol{U}^{i},\boldsymbol{U}^{i-1},\boldsymbol{\theta}\right) = \boldsymbol{0},\quad i=1,\cdots,N\nonumber\\
    &\quad\quad\ \boldsymbol{U}^0 = \boldsymbol{g}\left(\boldsymbol{\theta}\right).
\end{align}
where $\mathcal{L}:\mathbb{R}^{M\times N}\times\mathbb{R}^{L}\rightarrow\mathbb{R}$ is the objective function that reflects the system response. $\boldsymbol{\mathcal{U}}\in\mathbb{R}^{M\times N}$ is the stack of all state solutions with $\boldsymbol{U}^i$ as the solution at the $i$-th time step. The parameter vector $\boldsymbol{\theta}$ implicitly determines $\boldsymbol{U}^i$ through the residual function $\boldsymbol{R}^i$, and explicitly determines $\boldsymbol{U}^0$ through the initial value function $\boldsymbol{g}\left(\cdot\right):\mathbb{R}^L\rightarrow\mathbb{R}^M$. 

The total derivative of $\mathcal{L}$ with respect to $\boldsymbol{\theta}$ is necessary for solving (\ref{eq_inv}) with efficient gradient-based algorithms, which are difficult to obtain through analytical derivations.
In \texttt{DiffLiB}, we leverage the reverse-mode AD and define the customized VJP rule to compute $\frac{\textrm{d}\mathcal{L}}{\textrm{d}\boldsymbol{\theta}}$, which is non-trivial due to the implicit dependency of $\boldsymbol{U}^i$ on $\boldsymbol{\theta}$. 
For clarity, we first introduce the AD-based sensitivity formulation for $\mathcal{L}$ depending on a single-time response, and then extend it to a more general case where $\mathcal{L}$ relies on multi-time responses.

\subsubsection{Single-time response}
\label{Sec:inv:simp}
We first consider a objective function that only depends on the solution at the final time step $i=N$, i.e., $\mathcal{L}=\mathcal{L}(\boldsymbol{U}^N)$. The total derivative $\frac{\textrm{d}\mathcal{L}}{\textrm{d}\boldsymbol{\theta}}$ can be obtained from chain rules:
\begin{align}
\label{eq_inv_chain_single}
    \frac{\textrm{d}\mathcal{L}}{\textrm{d}\boldsymbol{\theta}} = \frac{\partial\mathcal{L}}{\partial\boldsymbol{\theta}} + \frac{\partial\mathcal{L}}{\partial\boldsymbol{U}^N}\frac{\textrm{d}\boldsymbol{U}^N}{\textrm{d}\boldsymbol{\theta}}
\end{align}
Since $\mathcal{L}$ is explicitly defined in terms of $\boldsymbol{U}^N$ and $\boldsymbol{\theta}$, the computation of the first two partial derivatives is straightforward. To apply the reverse-mode AD, we define a customized VJP rule for $\boldsymbol{U}^i$, enabling the computation of the implicit derivative $\frac{\textrm{d}\boldsymbol{U}^N}{\partial\boldsymbol{\theta}}$. Consider $\boldsymbol{U}^i$ as a function of $\boldsymbol{U}^{i-1}$ and $\boldsymbol{\theta}$, i.e., $\boldsymbol{U}^i=\boldsymbol{U}^i(\boldsymbol{U}^{i-1},\boldsymbol{\theta})$, the implicit function theorem \cite{MR0055409} gives:
\begin{align}
\label{eq_implict_single}
    \frac{\partial\boldsymbol{R}^i}{\partial\boldsymbol{U}^{i-1}} &+ \frac{\partial\boldsymbol{R}^i}{\partial\boldsymbol{U}^i}\frac{\partial\boldsymbol{U}^i}{\partial\boldsymbol{U}^{i-1}} = 0\nonumber\\
    \frac{\partial\boldsymbol{R}^i}{\partial\boldsymbol{\theta}} &+ \frac{\partial\boldsymbol{R}^i}{\partial\boldsymbol{U}^i}\frac{\partial\boldsymbol{U}^i}{\partial\boldsymbol{\theta}} = 0
\end{align}
Then we define the customized VJP rule for $\boldsymbol{U}^i$ as:
\begin{align}
\label{eq_vjp_single}
    \left(\boldsymbol{v}^i\right)^{\textrm{T}}\frac{\partial\boldsymbol{U}^i}{\partial\boldsymbol{U}^{i-1}} &=  \left(\boldsymbol{\lambda}^i\right)^{\textrm{T}}\frac{\partial\boldsymbol{R}^i}{\partial\boldsymbol{U}^{i-1}}\nonumber\\
    \left(\boldsymbol{v}^i\right)^{\textrm{T}}\frac{\partial\boldsymbol{U}^i}{\partial\boldsymbol{\theta}} &=  \left(\boldsymbol{\lambda}^i\right)^{\textrm{T}}\frac{\partial\boldsymbol{R}^i}{\partial\boldsymbol{\theta}}
\end{align}
where $\boldsymbol{v}^i$ is the propagated product vector. The adjoint vector $\boldsymbol{\lambda}^i$ can be obtained by solving the following linear adjoint equation:
\begin{align}
\label{eq_vjp2_single}
    \left(\frac{\partial\boldsymbol{R}^i}{\partial\boldsymbol{U}^i}\right)^{\textrm{T}}\boldsymbol{\lambda}^i = -\boldsymbol{v}^i
\end{align}

To derive the final expression for $\frac{\textrm{d}\mathcal{L}}{\textrm{d}\boldsymbol{\theta}}$ with the customized VJP rule, we employ the chain rules for $\boldsymbol{U}^i=\boldsymbol{U}^i(\boldsymbol{U}^{i-1},\boldsymbol{\theta})$:
\begin{align}
\label{eq_dU_single}
    \frac{\textrm{d}\boldsymbol{U}^i}{\textrm{d}\boldsymbol{\theta}} = \frac{\partial\boldsymbol{U}^{i}}{\partial\boldsymbol{\theta}} + \frac{\partial\boldsymbol{U}^{i}}{\partial\boldsymbol{U}^{i-1}}\frac{\textrm{d}\boldsymbol{U}^{i-1}}{\textrm{d}\boldsymbol{\theta}}
\end{align}
which can be submitted into (\ref{eq_inv_chain_single}) to obtain:
\begin{align}
\label{eq_inv_chain2_single}
   \frac{\textrm{d}\mathcal{L}}{\textrm{d}\boldsymbol{\theta}} = \frac{\partial\mathcal{L}}{\partial\boldsymbol{\theta}} + \frac{\partial\mathcal{L}}{\partial\boldsymbol{U}^N}\frac{\partial\boldsymbol{U}^{N}}{\partial\boldsymbol{\theta}} + \frac{\partial\mathcal{L}}{\partial\boldsymbol{U}^N}\frac{\partial\boldsymbol{U}^{N}}{\partial\boldsymbol{U}^{N-1}}\frac{\textrm{d}\boldsymbol{U}^{N-1}}{\textrm{d}\boldsymbol{\theta}}
\end{align}
Note that $\frac{\partial\mathcal{L}}{\partial\boldsymbol{U}^N}\frac{\partial\boldsymbol{U}^{N}}{\partial\boldsymbol{\theta}}$ and $\frac{\partial\mathcal{L}}{\partial\boldsymbol{U}^N}\frac{\partial\boldsymbol{U}^{N}}{\partial\boldsymbol{U}^{N-1}}$ have the VJP structure as shown in (\ref{eq_vjp_single}), we define $\boldsymbol{v}^N=\frac{\partial\mathcal{L}}{\partial\boldsymbol{U}^N}$ and obtain:
\begin{align}
\label{eq_inv_chain3_single}
      \frac{\textrm{d}\mathcal{L}}{\textrm{d}\boldsymbol{\theta}} = \frac{\partial\mathcal{L}}{\partial\boldsymbol{\theta}} + \boldsymbol{v}^N\frac{\partial\boldsymbol{U}^{N}}{\partial\boldsymbol{\theta}} + \boldsymbol{v}^N\frac{\partial\boldsymbol{U}^{N}}{\partial\boldsymbol{U}^{N-1}}\frac{\textrm{d}\boldsymbol{U}^{N-1}}{\textrm{d}\boldsymbol{\theta}}
\end{align}
which can be further expressed as:
\begin{align}
\label{eq_inv_chain4_single}
      \frac{\textrm{d}\mathcal{L}}{\textrm{d}\boldsymbol{\theta}} = \frac{\partial\mathcal{L}}{\partial\boldsymbol{\theta}} + \left(\boldsymbol{\lambda}^N\right)^{\textrm{T}}\frac{\partial\boldsymbol{R}^{N}}{\partial\boldsymbol{\theta}} + \left(\boldsymbol{\lambda}^N\right)^{\textrm{T}}\frac{\partial\boldsymbol{R}^{N}}{\partial\boldsymbol{U}^{N-1}}\frac{\textrm{d}\boldsymbol{U}^{N-1}}{\textrm{d}\boldsymbol{\theta}}
\end{align}
We observe that $(\boldsymbol{\lambda}^N)^{\textrm{T}}\frac{\partial\boldsymbol{R}^{N}}{\partial\boldsymbol{U}^{N-1}}\in\mathbb{R}^M$ is also a vector, which can be regarded as $\boldsymbol{v}^{N-1}$. Further expanding $\frac{\textrm{d}\boldsymbol{U}^{N-1}}{\textrm{d}\boldsymbol{\theta}}$ via (\ref{eq_dU_single}) yields new VJP structures: $\boldsymbol{v}^{N-1}\frac{\partial\boldsymbol{U}^{N-1}}{\partial\boldsymbol{\theta}}$ and $\boldsymbol{v}^{N-1}\frac{\partial\boldsymbol{U}^{N-1}}{\partial\boldsymbol{U}^{N-2}}$, along with a new total derivative $\frac{\textrm{d}\boldsymbol{U}^{N-2}}{\textrm{d}\boldsymbol{\theta}}$. This recursive expansion will continue until reaching $\frac{\textrm{d}\boldsymbol{U}^{0}}{\textrm{d}\boldsymbol{\theta}}$, and the propagated product vector in the emerging VJP structures can be summarized as:
\begin{align}
\label{eq_vjp3_single}
    \left(\boldsymbol{v}^N\right)^{\textrm{T}}&=\frac{\partial\mathcal{L}}{\partial\boldsymbol{U}^N}\nonumber\\
    \left(\boldsymbol{v}^i\right)^{\textrm{T}} &= \left(\boldsymbol{\lambda}^{i+1}\right)^{\textrm{T}}\frac{\partial\boldsymbol{R}^{i+1}}{\partial\boldsymbol{U}^i},\quad i=N-1,\cdots,0
\end{align}
Since the initial solution $\boldsymbol{U}^{0}$ is determined by the explicit function $\boldsymbol{g}$ of $\boldsymbol{\theta}$, no implicit derivatives will be involved in the final expression for $\frac{\textrm{d}\mathcal{L}}{\textrm{d}\boldsymbol{\theta}}$, which can be expressed as:
\begin{align}\label{eq_inv_chain5_single}
        \frac{\textrm{d}\mathcal{L}}{\textrm{d}\boldsymbol{\theta}} = \frac{\partial\mathcal{L}}{\partial\boldsymbol{\boldsymbol{\theta}}} + \sum_{i=1}^N\left(\boldsymbol{\lambda}^i\right)^{\textrm{T}}\frac{\partial\boldsymbol{R}^i}{\partial\boldsymbol{\theta}}+\left(\boldsymbol{\lambda}^0\right)^{\textrm{T}}\frac{\textrm{d}\boldsymbol{g}}{\textrm{d}\boldsymbol{\theta}}
\end{align}
where $\boldsymbol{\lambda}^0=\left(\frac{\partial\boldsymbol{R}^1}{\partial\boldsymbol{U}^0}\right)^{\textrm{T}}\boldsymbol{\lambda}^1$. All the derivatives and adjoint vectors in (\ref{eq_inv_chain5_single}) are automatically computed via the customized reverse-mode AD, eliminating the need for problem-specific derivations. 

\begin{figure}[H] \centering
    {\includegraphics[width=1\textwidth]{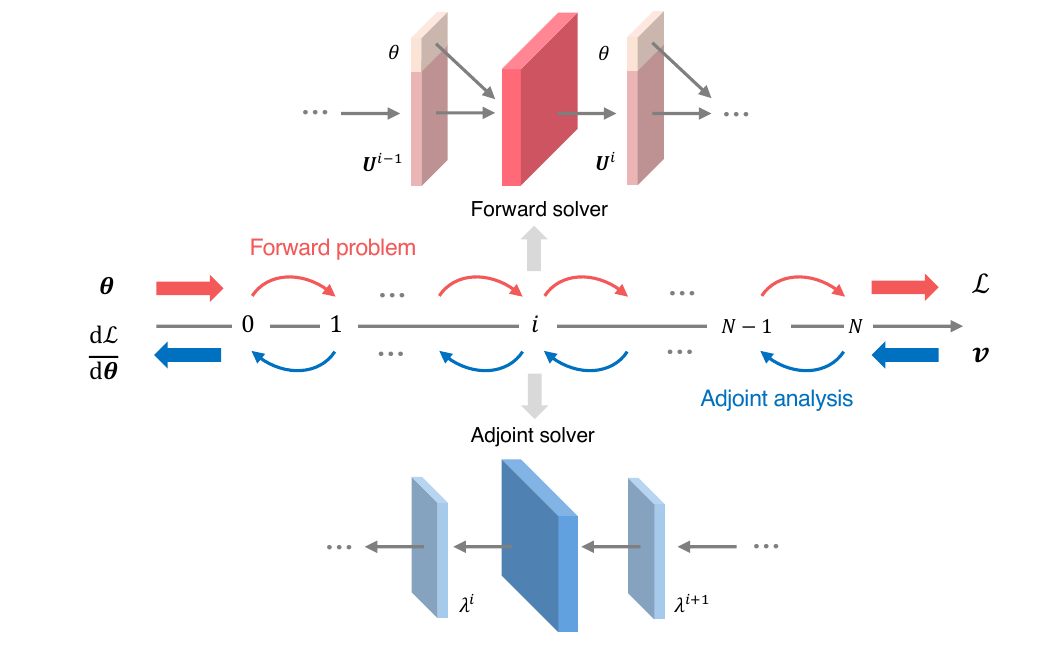}}
    \caption{The workflow of the AD-based sensitivity analysis framework.} \label{fig:ad_simp}
\end{figure}

The complete workflow is summarized in Fig.~\ref{fig:ad_simp}: The computation begins with solving the forward problem. At the $i$-th time step, the parameter $\boldsymbol{\theta}$ and the previous solution $\boldsymbol{U}^{i-1}$ are input to the forward solver to compute $\boldsymbol{U}^i$, which will be stored in memory for subsequent computations. After the evaluation of $\mathcal{L}$, the reverse propagation is initiated with a seed vector and propagates from the output. For each time step, the adjoint vector $\boldsymbol{\lambda}^i$ is obtained through the adjoint solver (\ref{eq_vjp2_single}) and then propagated towards the input. Consequently, the reverse propagation involves solving $N$ adjoint equations, each of the same size as (\ref{eq_mac_inc}) in the forward problem. Since solving linear systems dominates the computational cost, the evaluation of $\frac{\textrm{d}\mathcal{L}}{\textrm{d}\boldsymbol{\theta}}$ requires approximately twice the effort of solving the forward problem alone.

\subsubsection{Multi-time responses}
A more general case involves considering the responses at multiple time steps in the objective function, such as the variation of the terminal voltage over time. The expression for $\frac{\textrm{d}\mathcal{L}}{\textrm{d}\boldsymbol{\theta}}$ becomes:
\begin{align}
\label{eq_inv_chain_multi}
    \frac{\textrm{d}\mathcal{L}}{\textrm{d}\boldsymbol{\theta}} = \frac{\partial\mathcal{L}}{\partial\boldsymbol{\theta}} + \sum_{i=1}^{N}\frac{\partial\mathcal{L}}{\partial\boldsymbol{U}^i}\frac{\textrm{d}\boldsymbol{U}^i}{\textrm{d}\boldsymbol{\theta}}
\end{align}
where all the implicit derivatives $\frac{\textrm{d}\boldsymbol{U}^i}{\textrm{d}\boldsymbol{\theta}}$ can also be handled by the AD-based sensitivity analysis framework. It should be noted that the summation in (\ref{eq_inv_chain_multi}) is not performed sequentially but also obtained in one reverse propagation. The only difference compared to the single-time response is that the propagated product vector $\boldsymbol{v}^i$ consists of two parts: the contribution from the output and the upstream solution, which can be expressed as:
\begin{align}
\label{eq_vjp_multi}
    \left(\boldsymbol{v}^i\right)^{\textrm{T}} = \frac{\partial\mathcal{L}}{\partial\boldsymbol{U}^i}+\left(\boldsymbol{\lambda}^{i+1}\right)^{\textrm{T}}\frac{\partial\boldsymbol{R}^{i+1}}{\partial\boldsymbol{U}^i},\quad i=N-1,\cdots,0
\end{align}
with $\boldsymbol{v}^N$ remaining the same as (\ref{eq_vjp3_single}). This change arises because $\boldsymbol{U}^i$ serves both as a component of $\mathcal{L}$ and an intermediate variable for the upstream solution, resulting in two different propagation paths in the computational graph. 

In summary, the cornerstone of the AD-based sensitivity analysis lies in defining the customized VJP rule (\ref{eq_vjp_single}) through the adjoint-based implicit differentiation method \cite{blondel2022efficient}, enabling the evaluation of implicit derivatives in the chain rule. The remaining operations, such as the recursive propagation (\ref{eq_vjp_multi}), are processed automatically by the robust AD framework in \texttt{JAX} without manual intervention. In addition, we present in Appendix \ref{app_adj} the adjoint sensitivity analysis of (\ref{eq_inv}) via the Lagrange multiplier method \cite{cea1986conception}. The derivation yields the same results for both single-time and multi-time responses, providing further validation for the AD-based sensitivity analysis framework.

\section{Numerical examples}
\label{Sec:examples}
This section gives several numerical examples to demonstrate the accuracy and effectiveness of \texttt{DiffLiB} in solving both forward and inverse problems, which includes:
\begin{enumerate}
    \item 2D forward predictions benchmarked with \texttt{COMSOL};
    \item 2D inverse parameter identification with experiment data; 
    \item 3D forward predictions benchmarked with \texttt{COMSOL};
    \item 3D inverse parameter identification with simulation data.
\end{enumerate}
The first and third examples are designed to demonstrate \texttt{DiffLiB}'s adaptability to 2D and 3D battery geometries, i.e., the P3D and P4D models. The simulation results from the commercial software \texttt{COMSOL} are used as the ground truth for comparison to validate the accuracy of \texttt{DiffLiB} in forward predictions. The second example involves identifying model parameters based on experiment data from C-rate tests, aligning with practical workflows in the battery industry. The final example addresses a 3D parameter identification problem with simulation data, highlighting \texttt{DiffLiB}'s capability to solve inverse problems with real battery geometries.

\subsection{2D benchmark problems}
\label{Sec:example:bmk}
In the first example, we consider the constant-current discharge of a LIB cell over a range of current rates. 
The model parameters and functions (e.g., $U_{\textrm{oc}}$) are taken from \cite{marquis2019asymptotic} and given in Appendix~\ref{app_para_Ma2019}, which corresponds to a LIB cell with a graphite anode and a lithium-cobalt oxide cathode. 
Five different rates, $0.2\rm~C$, $0.5\rm~C$, $1.0\rm~C$, $1.5\rm~C$, and $2.0\rm~C$, are considered here to verify the accuracy of \texttt{DiffLiB} in forward predictions, where the discharging current of $1\rm~C$ corresponds to $24\rm~A/m^2$. 
The discharging process is terminated if the terminal voltage reach its lower limit $3.105\rm~V$.

The macroscopic computational domain is set as a $225\times 10$ rectangle, which is discretized with $60$ bilinear quadrilateral elements. The microscopic computational domain is defined at each mesh node within the electrode subdomain and discretized with $10$ linear interval elements. The same settings are used in \texttt{COMSOL} to generate the reference data.

\begin{figure}[H] \centering
    \subfigure[]
    {\includegraphics[width=0.47\textwidth]{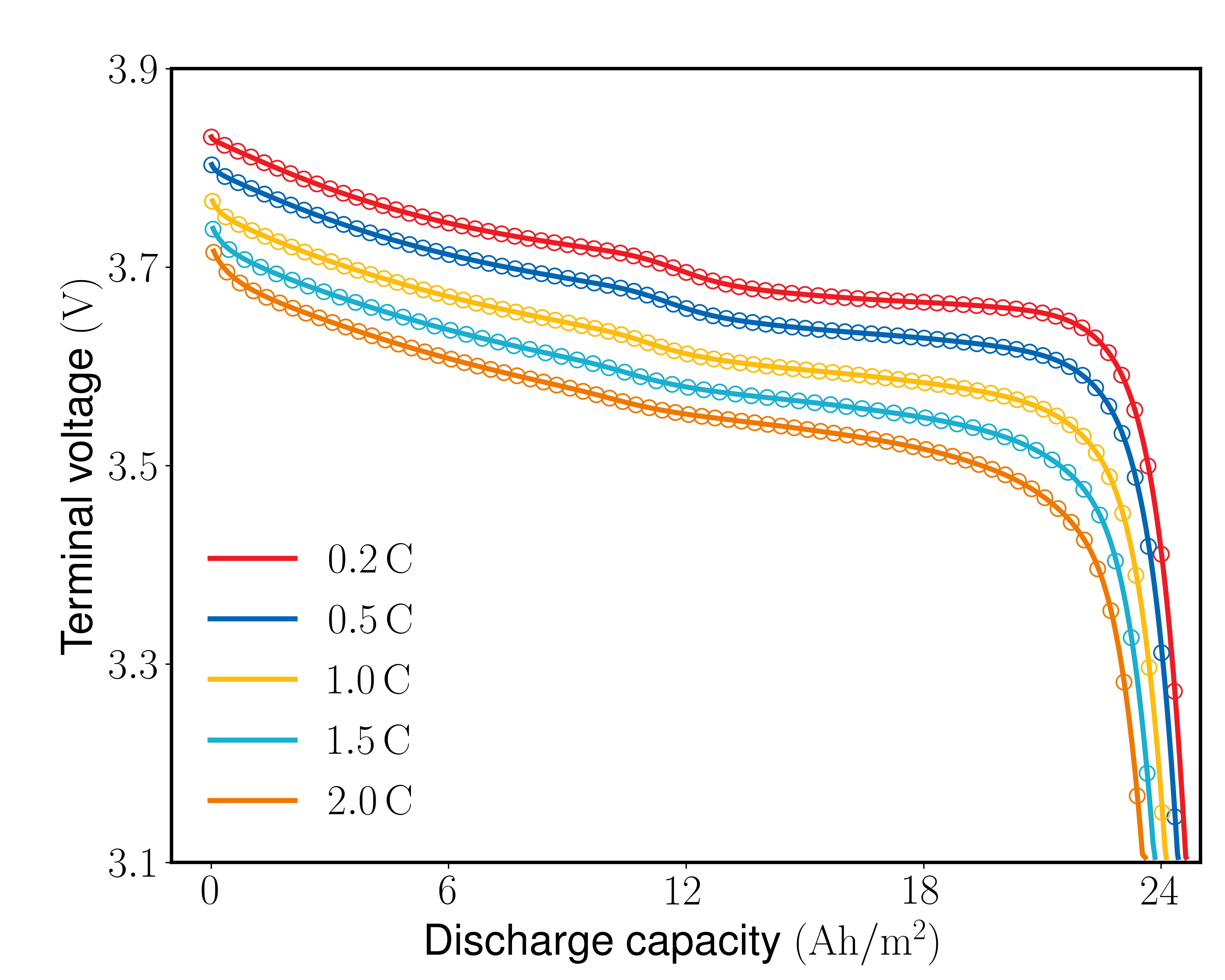}\label{fig:bmk_v_a}}
    \hspace{0.01\textwidth}
    \subfigure[]
    {\includegraphics[width=0.47\textwidth]{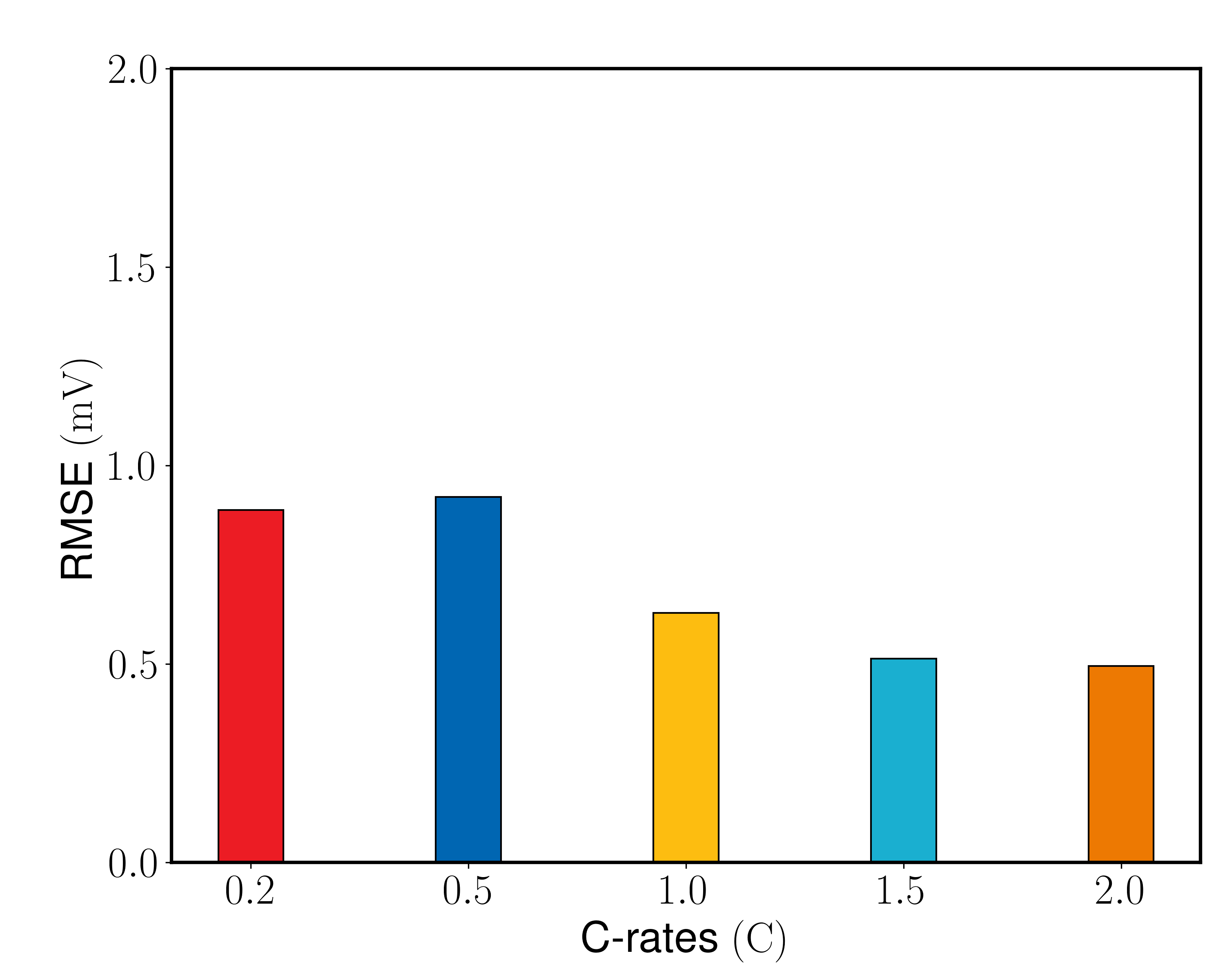}\label{fig:bmk_v_b}} 
    \caption{2D validation of the terminal voltage under different discharging currents. (Solid lines: \texttt{DiffLiB}; markers: \texttt{COMSOL}.) (a) Variation of the terminal voltage during the discharging process. (b) Root mean square errors (RMSE) of the terminal voltage. } \label{fig:bmk_v}
\end{figure}

\begin{figure}[H] \centering
    \subfigure[]
    {\includegraphics[width=0.47\textwidth]{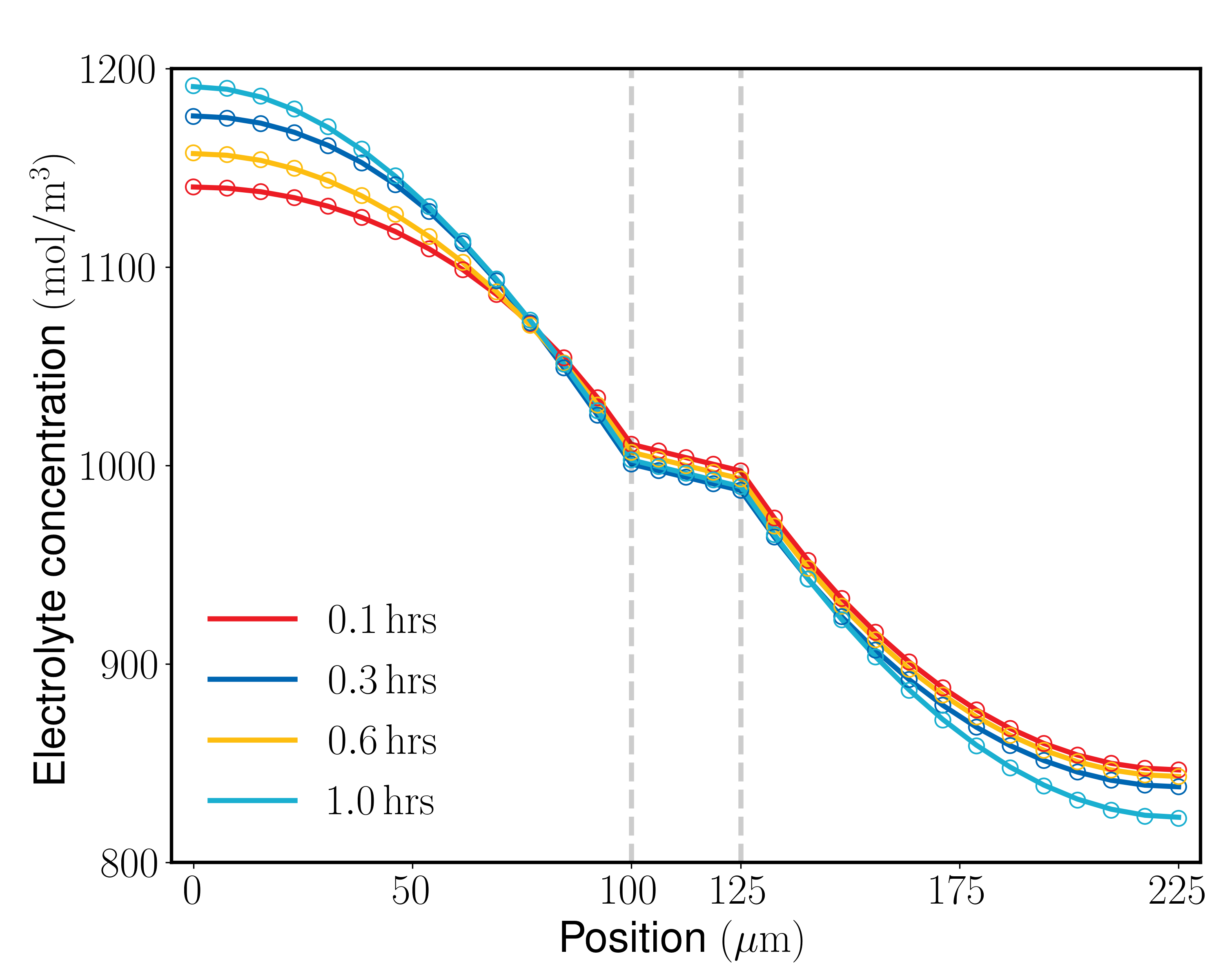}\label{fig:bmk_var_a}}
    \hspace{0.01\textwidth}
    \subfigure[]
    {\includegraphics[width=0.47\textwidth]{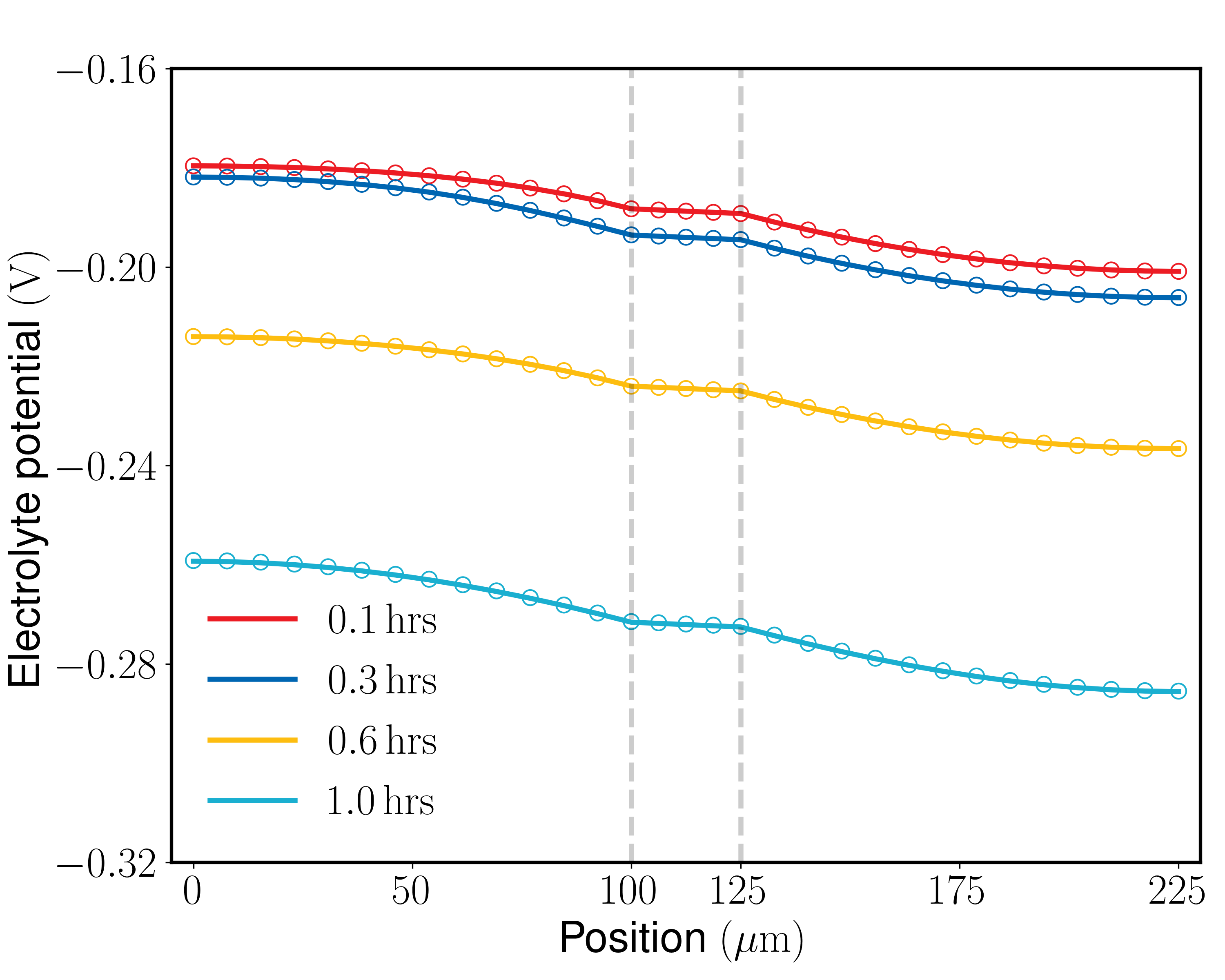}\label{fig:bmk_var_b}} 
    \\
    \subfigure[]
    {\includegraphics[width=0.47\textwidth]{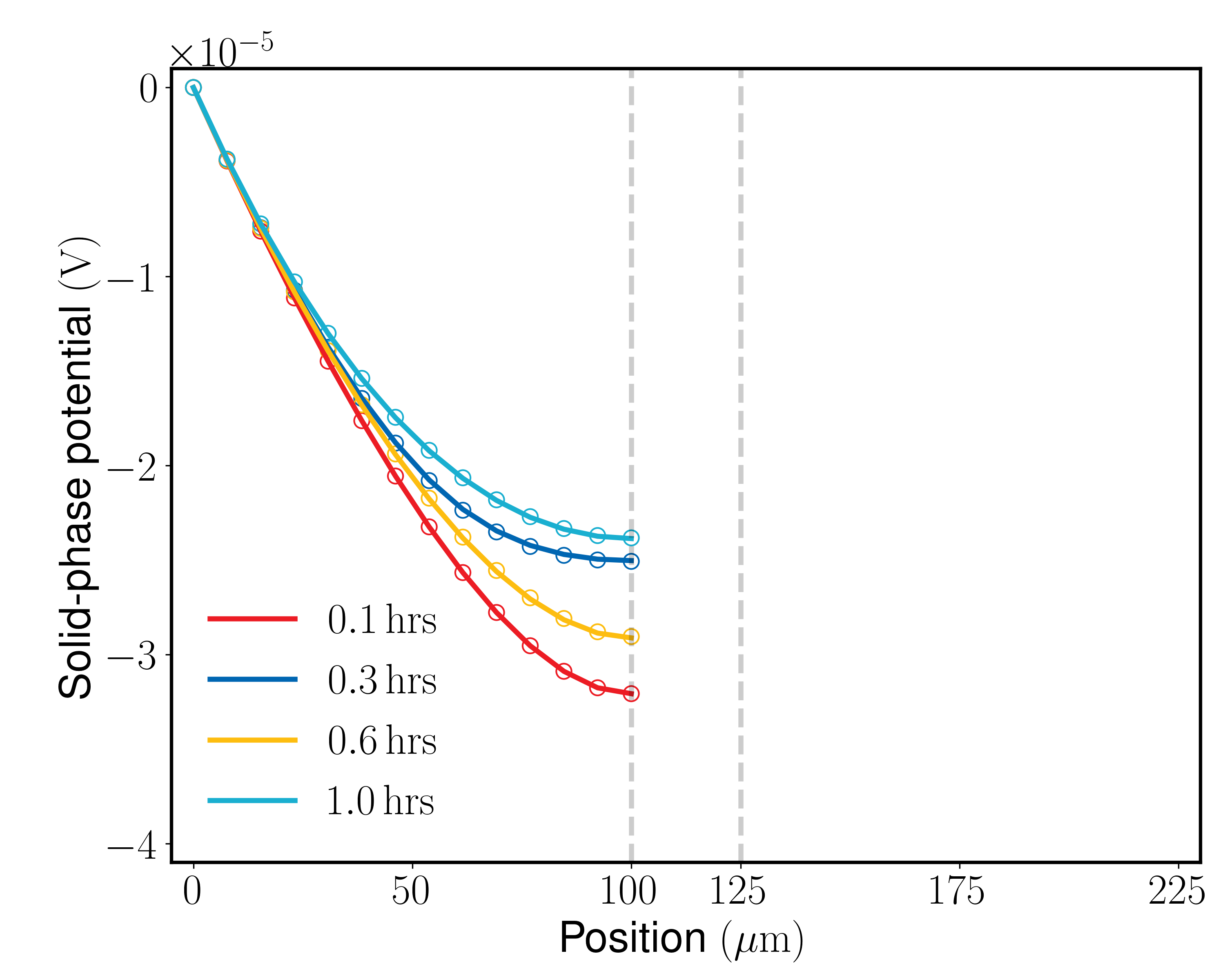}\label{fig:bmk_var_c}} 
    \hspace{0.01\textwidth}
    \subfigure[]
    {\includegraphics[width=0.47\textwidth]{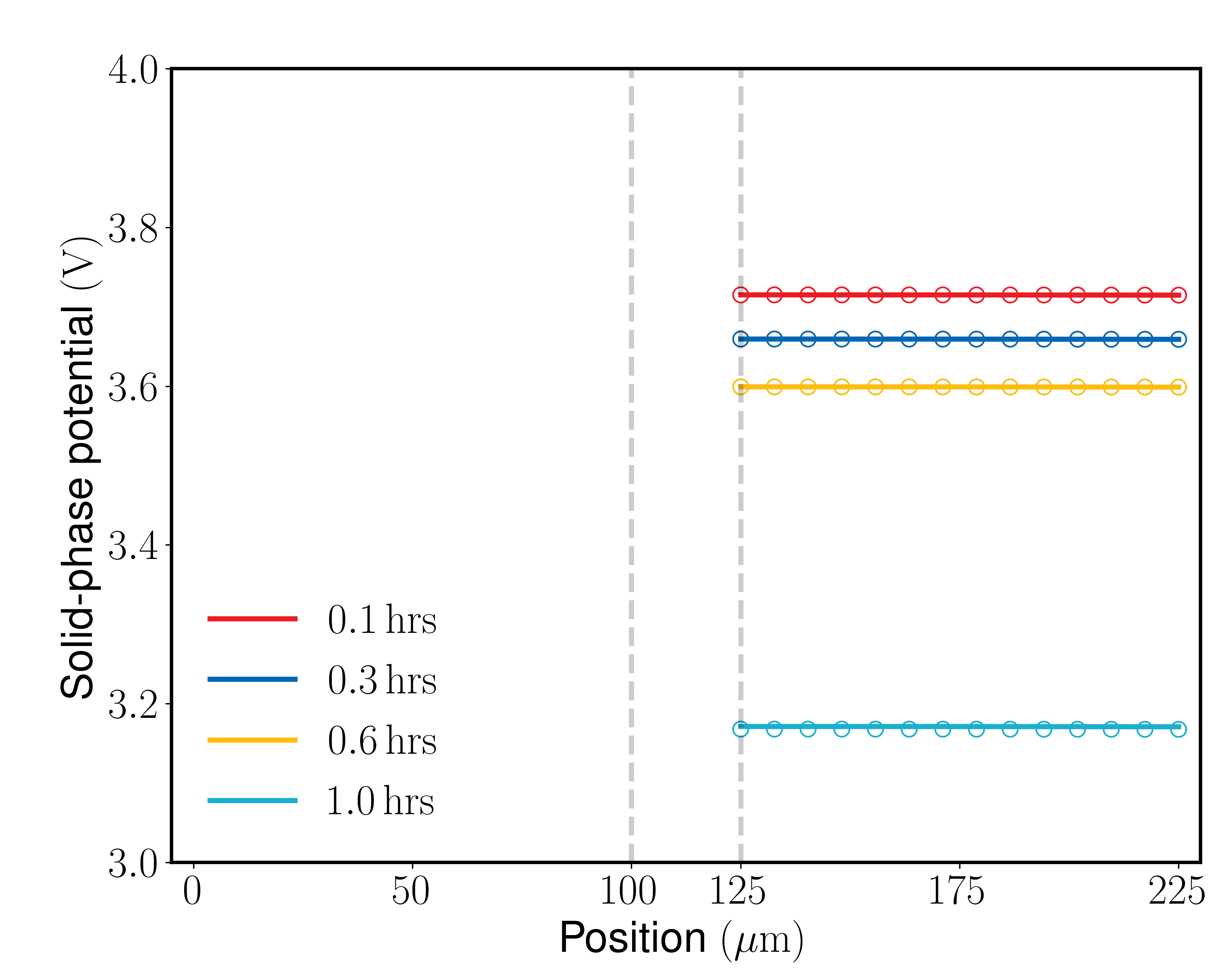}\label{fig:bmk_var_d}} 
    \caption{2D validation of the state variable distributions under the discharging current of $1\rm~C$. (Solid lines: \texttt{DiffLiB}; markers: \texttt{COMSOL}.) (a) Electrolyte lithium concentration. (b) Electrolyte potential. (c) Anode solid-phase potential. (d) Cathode solid-phase potential.} \label{fig:bmk_var}
\end{figure}

The comparison results are presented in Figs.~(\ref{fig:bmk_v}) and (\ref{fig:bmk_var}). The solid lines represent the results from \texttt{DiffLiB}, and markers denote the reference data obtained from \texttt{COMSOL}, serving as the ground truth. Fig.~\ref{fig:bmk_v_a} shows the variation of terminal voltage during the discharging process under different current rates. We observe that the prediction results from \texttt{DiffLiB} closely align with those of \texttt{COMSOL} under both low and high current rates. The root mean square errors (RMSE) of the terminal voltage are also calculated and presented in Fig.~\ref{fig:bmk_v_b}. All RMSE values remain below $1\rm~mV$, which is less than $0.1\%$ of the predicted values, validating the agreement between \texttt{DiffLiB} and \texttt{COMSOL} in forward predictions.

Figs.~\ref{fig:bmk_var_a} to \ref{fig:bmk_var_d} illustrate the distributions of state variables along the cell thickness under the discharging current of $1\rm~C$, including the electrolyte lithium concentration $c_{\textrm{e}}$, the electrolyte potential $\phi_e$, and the solid-phase potential $\phi_s$ in the anode and cathode. The region between the two gray dashed lines represents the separator, with the anode and cathode located on the left and right sides, respectively. We observe that the distributions of state variables obtained from \texttt{DiffLiB} demonstrate high consistency with those from \texttt{COMSOL} across different periods of the discharging process, further validating the accuracy of \texttt{DiffLiB} in forward predictions.

\subsection{Parameter identification with experiment data}
\label{Sec:examples:exp}
Accurate parameterization is critical for ensuring the predictive accuracy and practical utility of battery simulation models, which is a long-standing challenge in the battery industry. A key advantage of \texttt{DiffLiB} is its ability to automatically compute sensitivity information for input parameters, enabling efficient solutions to inverse problems such as parameter identification. This subsection addresses a parameter identification problem for a commercial LIB cell using experiment data, demonstrating the potential of \texttt{DiffLiB} in industrial applications.

The model parameters and functions of a commercial $\rm LiCoO_2$-Graphite battery cell are shown in Appendix \ref{app_exp_para}, some of which have been determined by the manufacturing information. Eight parameters (marked with TBI in Table~\ref{tab:para_exp}) remain to be identified, which can be further classified into two categories:
\begin{enumerate}
    \item Thermodynamic parameters: the initial microscopic lithium concentrations of the anode ($c_{\textrm{s,a}}^0$) and cathode ($c_{\textrm{s,c}}^0$). 
    \item Kinetic parameters: the Bruggeman coefficients for the anode ($\beta_\textrm{a}$) , separator($\beta_{\textrm{se}}$), and cathode ($\beta_\textrm{c}$); the transference number $t_+$; the reaction rate constants for the anode ($k_\textrm{s,a}$) and cathode ($k_\textrm{s,c}$).
\end{enumerate}

\begin{figure}[H] \centering
    {\includegraphics[width=0.6\textwidth]{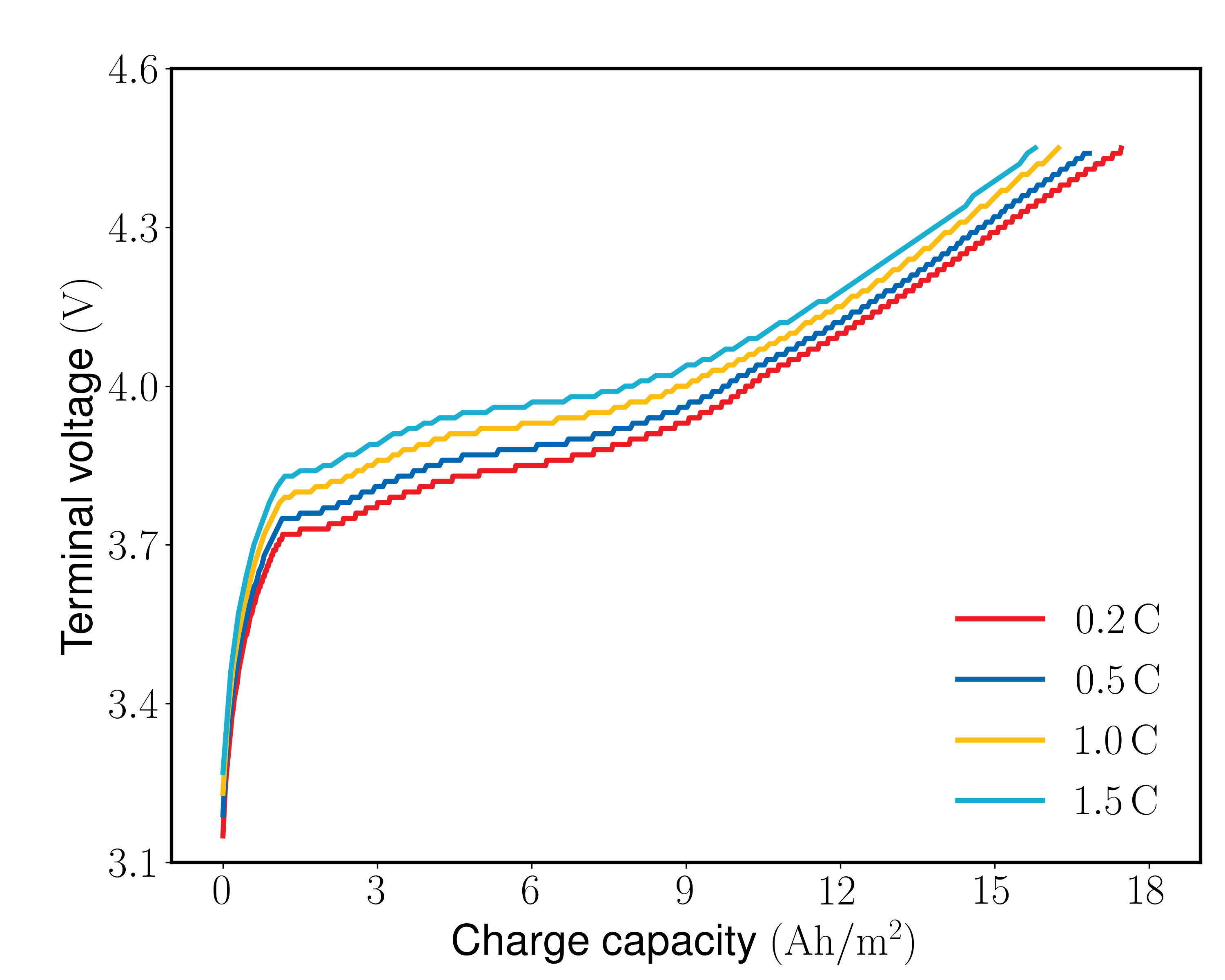}}
    \caption{Reference terminal voltage from experimental C-rate tests of a commercial LIB cell.} \label{fig:exp_data}
\end{figure}

Experimental characterization is a critical step in the parameter identification for LIBs, as it provides reference data to calibrate and validate model parameters. 
A widely adopted approach for experimental characterization involves conducting C-rate tests, which evaluate LIB performance under different current loads \cite{laue2021practical}. 
Fig.~\ref{fig:exp_data} presents the measured terminal voltage of the commercial LIB cell during constant-current charge at four different rates ($0.2\rm~C$, $0.5\rm~C$, $1.0\rm~C$, and $1.5\rm~C$), where the charging current of $1\rm~C$ corresponds to $18.04\rm~A/m^2$. Each charging process is terminated after reaching the upper limit $4.45\rm~V$.

To enhance the adaptability of model parameters across different current loads, the objective function in (\ref{eq_inv}) is formulated as:
\begin{align}
\label{eq_para_inv}
    \mathcal{L} = \frac{1}{M}\sum_{i=1}^M\sqrt{\frac{1}{N}\sum_{j=1}^{N}\left(V_{i,j}\left(\boldsymbol{\theta}\right)-V_{i,j}^{\textrm{ref}}\right)^2}
\end{align}
where $V_{i,j}$ is the predicted terminal voltage at the $j$-th time step under the $i$-th charging current. 
$V_{i,j}^{\textrm{ref}}$ is the corresponding reference value obtained from the C-rate tests. 
Eq.~(\ref{eq_para_inv}) can be regarded as the average RMSE between the predicted and reference terminal voltage under different charging currents. 
A smaller value of (\ref{eq_para_inv}) represents the better adaptability of the parameters, leading to higher reliability of the battery simulation model. 

The range of each parameter is obtained by referring to the previous research~\cite{jin2018parameter,laue2021practical,xu2022enabling}, which is shown in the second column of Table~\ref{tab:vars_exp}. Since the parameter range varies across different orders, we normalized the design variable using the scaling formulation of Jin et al.~\cite{jin2018parameter}, where $k_{\textrm{s,a}}$ and $k_{\textrm{s,c}}$ are scaled using:
\begin{align}
\label{eq_para_int1}
    \ln\boldsymbol{\theta}_i = w_i\ln\left(\underline{\boldsymbol{\theta}_i}\right)+\left(1-w_i\right)\ln\left(\overline{\boldsymbol{\theta}_i}\right)
\end{align}
The remaining parameters are scaled using:
\begin{align}
\label{eq_para_int2}
    \boldsymbol{\theta}_i = w_i\underline{\boldsymbol{\theta}_i}+\left(1-w_i\right)\overline{\boldsymbol{\theta}_i}
\end{align}
where $w_i\in\left[0,1\right]$ denotes the design variable of the $i$-th parameter $\boldsymbol{\theta}_i$. $\underline{\boldsymbol{\theta}_i}$ and $\overline{\boldsymbol{\theta}_i}$ are the lower and upper bound of $\boldsymbol{\theta}_i$. 

For comparison, the parameter identification problem is solved with two different optimizers: the limited-memory BFGS algorithm (L-BFGS-B) \cite{byrd1995limited} in \texttt{SciPy} \cite{virtanen2020scipy} and the strengthen elitist genetic algorithm (SEGA) in \texttt{Geatpy} \cite{jazzbin2020geatpy}. 

The gradient-based L-BFGS-B group (denoted as GB) starts from a initial value and updates the parameters using the sensitivity information from \texttt{DiffLiB}. The choice of the initial value is critical to the convergence and stability of the optimization process. Previous study \cite{laue2021practical} indicates that low current rates are suitable for identifying thermodynamic parameters, while high current rates are appropriate for kinetic parameters. Therefore, we employ a two-stage strategy to search for suitable initial values before conducting optimization with complete experiment data. Specifically, we initialize $w_i$ in (\ref{eq_para_int1}) and (\ref{eq_para_int2}) to $0.5$, followed by:
\begin{enumerate}
    \item Fix the kinetic parameters and optimize the thermodynamic parameters through (\ref{eq_para_inv}) using the experiment data at $0.2\rm~C$.
    \item Fix the obtained thermodynamic parameters and optimize the kinetic parameters through (\ref{eq_para_inv}) using the experiment data at $0.5\rm~C$, $1.0\rm~C$, and $1.5\rm~C$.  
\end{enumerate}
The obtained parameter combination serve as the initial value for the complete-data optimization. Since we focus on appropriate initial values rather than optimal solutions, the maximum optimization iteration in each stage is set to $10$, and each optimization process is terminated if the objective function is less than $10\rm~mV$.

The gradient-free SEGA first generates a population with several individuals, i.e., parameter combinations, and evaluates the objective function for each individual to determine the evolution of the population. To account for the randomness in the initial population, we consider three SEGA groups (denoted as GF(I), GF(II), and GF(III)) with the same population size of $64$ and different random seeds.

We utilize $90$ bilinear quadrilateral elements in the macroscopic computational domain while adopting $10$ linear interval elements for each microscopic computational domain. All computations are preformed on an Intel(R) Xeon(R) Platinum platform with $16$ cores and $32$ GB RAM. If the average of the absolute differences between the current objective function value and that from the previous nine iterations is less than $0.09\rm~mV$, the optimization is deemed to have stagnated and is terminated.

\begin{figure}[htbp] \centering
    {\includegraphics[width=0.55\textwidth]{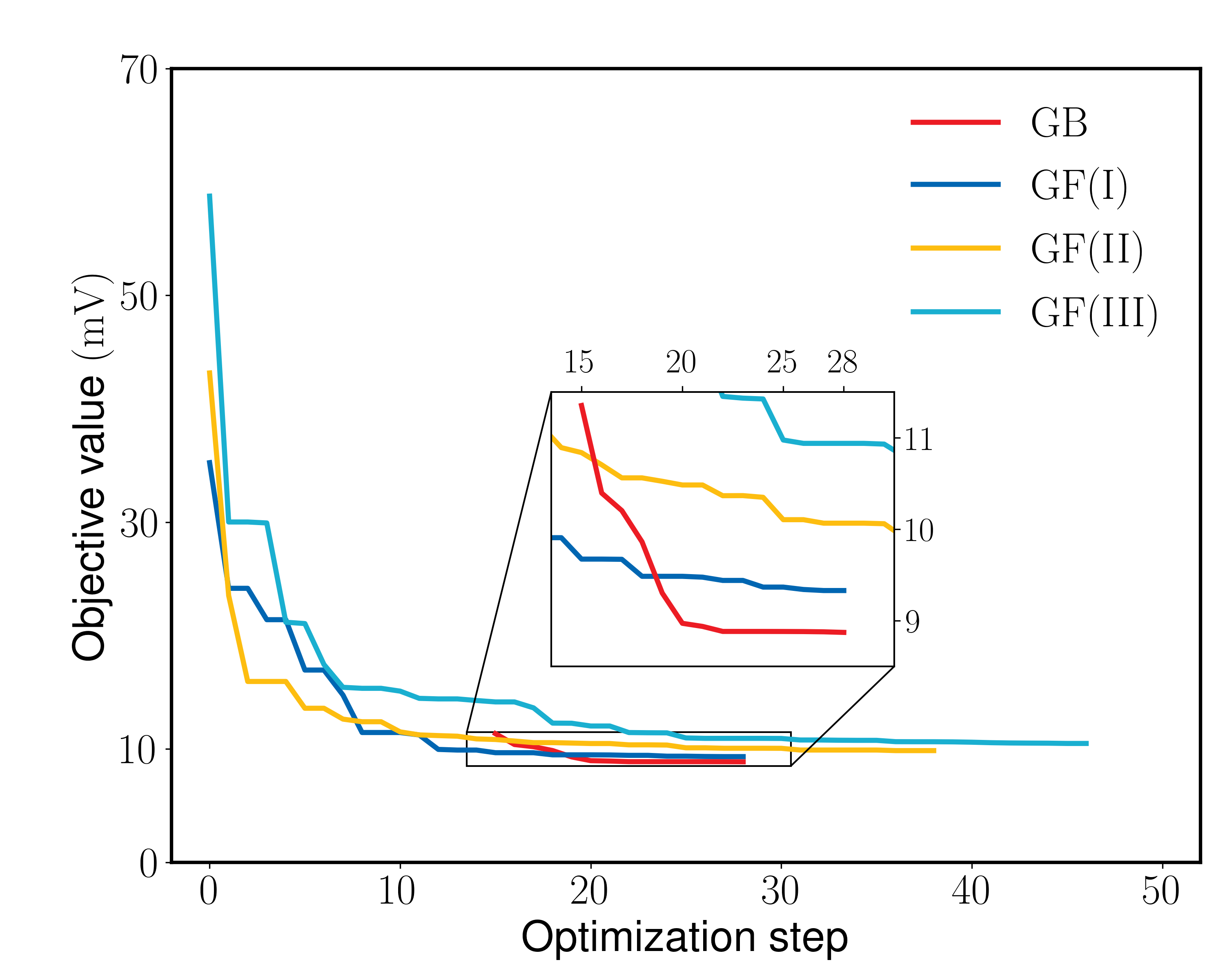}}
    \caption{Optimization history of the L-BFGS-B (GB) and SEGA (GF).} \label{fig:exp_his}
\end{figure}

\begin{figure}[H] \centering
    \subfigure[]
    {\includegraphics[width=0.47\textwidth]{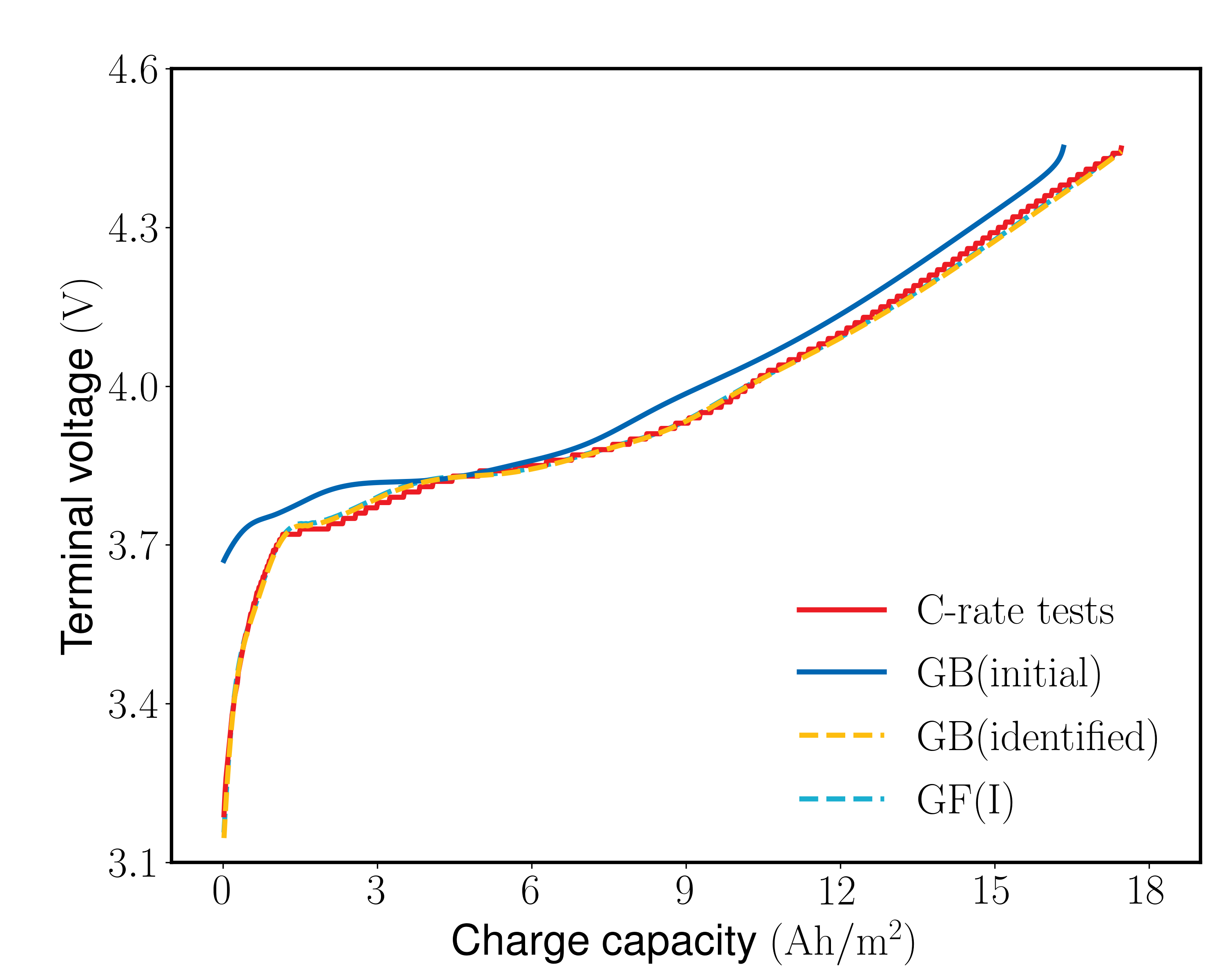}}
    \hspace{0.01\textwidth}
    \subfigure[]
    {\includegraphics[width=0.47\textwidth]{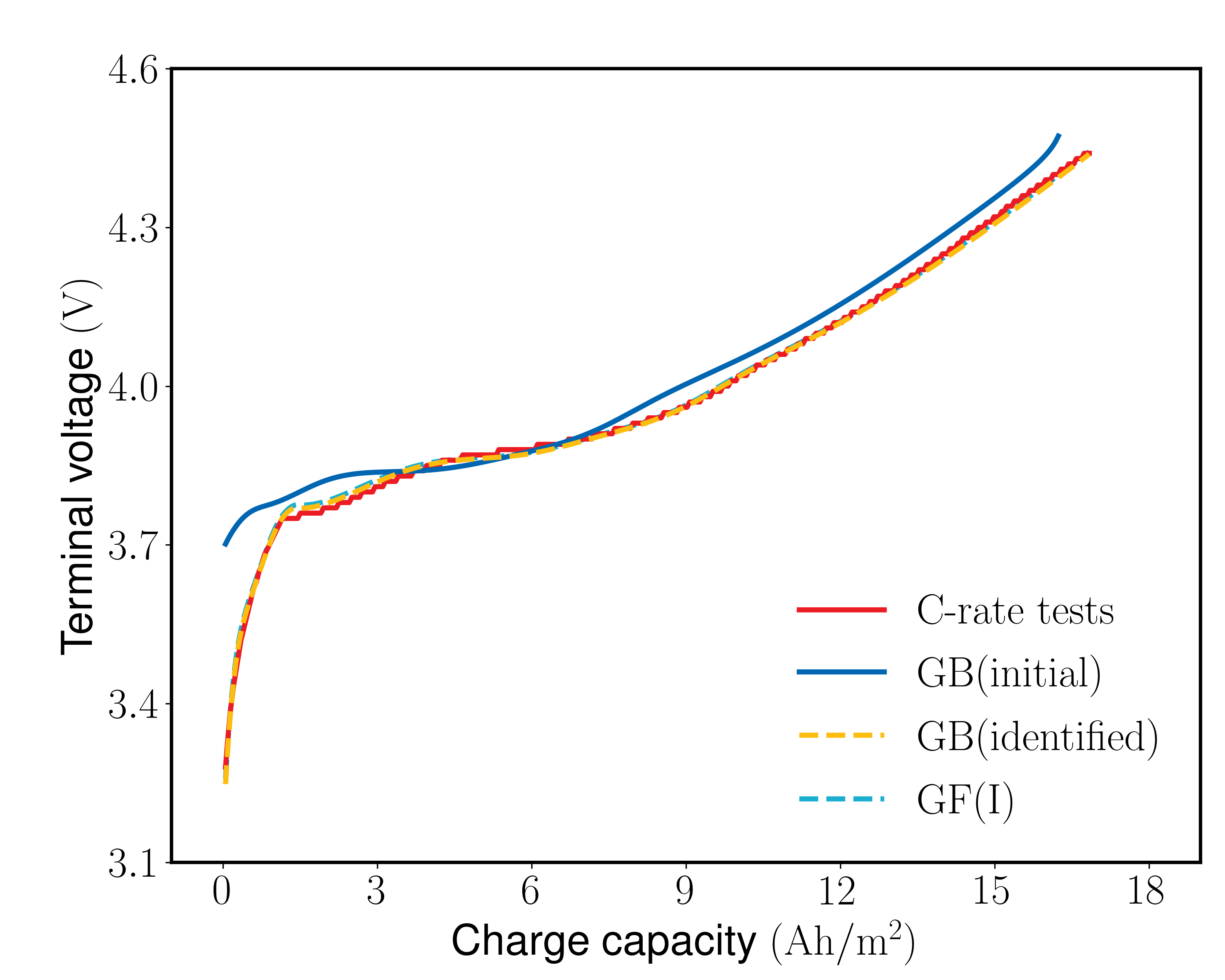}} 
    \\
    \subfigure[]
    {\includegraphics[width=0.47\textwidth]{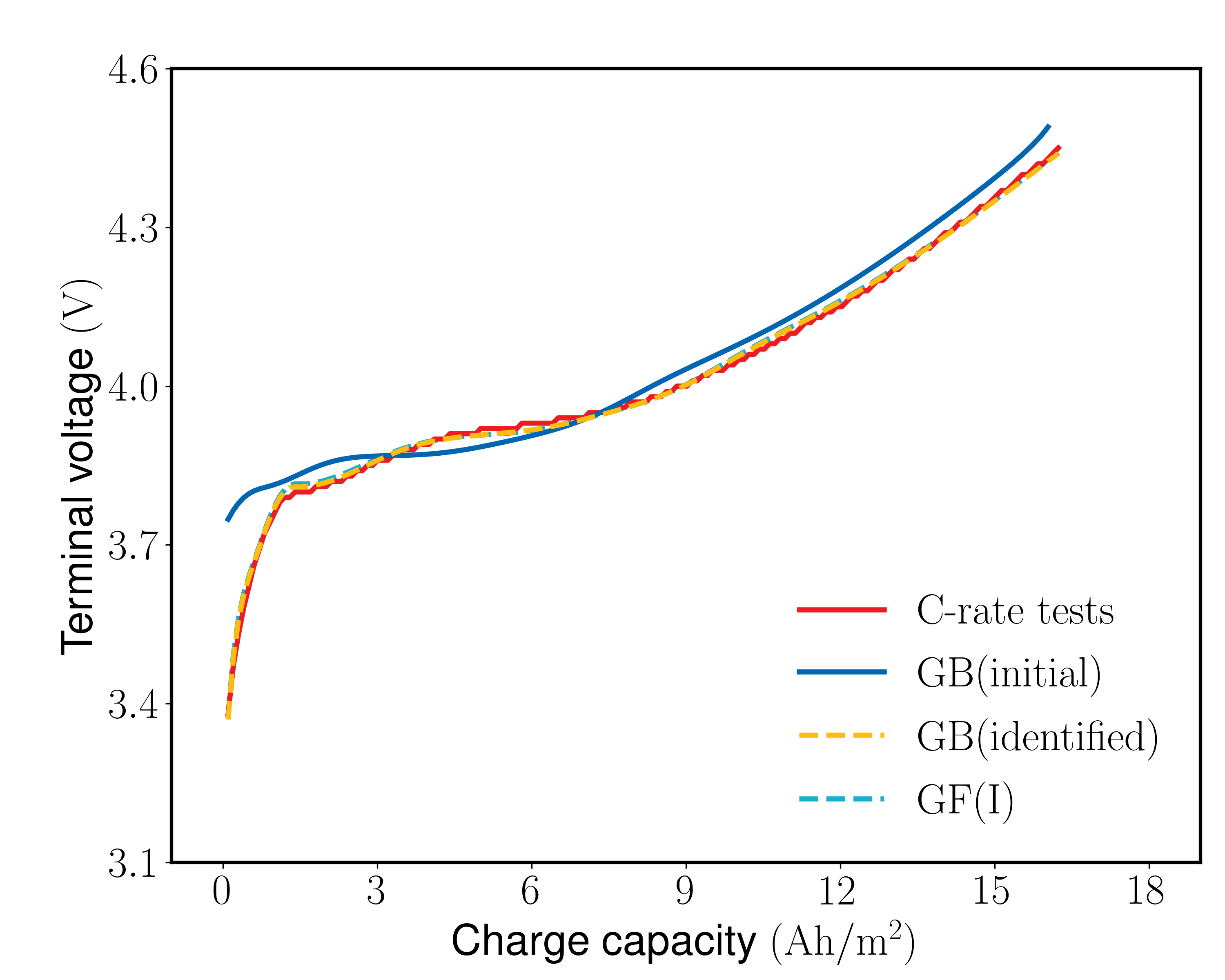}}
    \hspace{0.01\textwidth}
    \subfigure[]
    {\includegraphics[width=0.47\textwidth]{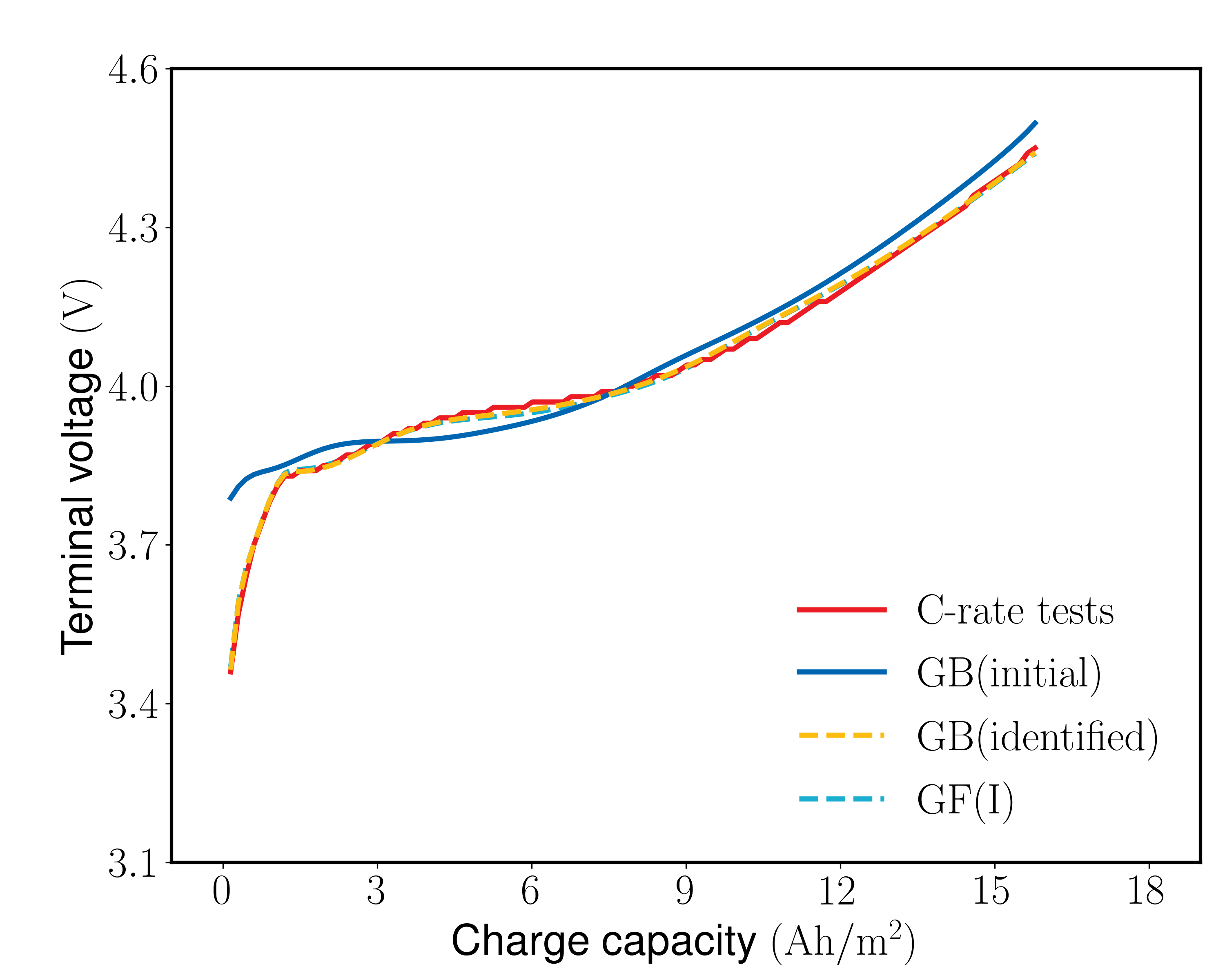}} 
    \caption{Validation of the identified parameters of L-BFGS-B (GB) and SEGA (GF). (a) $0.2\rm~C$. (b) $0.5\rm~C$. (c) $1.0\rm~C$. (d) $1.5\rm~C$.} \label{fig:exp_vad}
\end{figure}

The identified parameter values are presented in Table~\ref{tab:vars_exp}. 
To further analyze the optimization process, Fig.~\ref{fig:exp_his} shows the evolution of the objective function value across iterations. The GB achieves the best performance, attaining the minimal objective function value in only 28 iterations (with the first 15 iterations dedicated to initial value exploration).
Among the GF groups, GF(I) reaches the optimal value in 28 iterations, outperforming GF(II) (38 iterations) and GF(III) (46 iterations). 
Fig.~\ref{fig:exp_vad} compares the terminal voltage predicted by the identified parameters of GB and GF(I) with experimental measurements. The results obtained using the initial GB parameters ($w_i=0.5$) are also included for reference.
We observe that the initial GB parameters yield terminal voltage predictions inconsistent with experimental measurements. However, following the initial-value search strategy and gradient-based optimization with the complete experiment data, the identified GB parameters closely match the experimental measurements across different current loads, performing better than GF(I). This demonstrates the effectiveness of the implemented parameter identification scheme.

\begin{table}[htbp]
\centering
\caption{Optimization results of parameter identification through L-BFGS-B (GB) and SEGA (GF). (Obj: Objective function value. NF: number of forward predictions)}
\label{tab:res_exp}
\setlength{\tabcolsep}{4pt} 
\begin{tabularx}{\textwidth}{@{\hspace{5pt}} 
    >{\raggedright}X 
    >{\centering}X 
    *{4}{>{\centering\arraybackslash}X} 
    >{\centering\arraybackslash}X 
    >{\centering\arraybackslash}X 
    @{}}
\toprule
\multirow{2}{*}{\centering Group} & 
\multirow{2}{*}{\centering Obj ($\rm mV$)} & 
\multicolumn{4}{c}{RMSE ($\rm mV$)} & 
\multirow{2}{*}{\centering NF} & 
\multirow{2}{*}{\centering Time (h)} \\
\cmidrule(lr){3-6} 
& & $0.2\rm C$ & $0.5\rm C$ & $1.0\rm C$ & $1.5\rm C$ & & \\
\midrule
GB & 8.87 & 10.54 & 8.45 & 8.00 & 8.49 & \textbf{250} & \textbf{3.17} \\
GF(I) & 9.33 & 9.82 & 8.76 & 9.20 & 9.54 & 7424 & 11.64 \\
GF(II) & 9.86 & 11.15 & 9.64 & 9.11 & 9.54 & 9984 & 15.30 \\
GF(III) & 10.49 & 11.21 & 10.60 & 9.71 & 10.43 & 12032 & 18.74 \\
\bottomrule
\end{tabularx}
\end{table}

Table~\ref{tab:res_exp} summarizes the optimization results for all groups, including the final objective function values (Obj), the RMSE between the predicted and reference terminal voltage for each current load, the number of forward predictions (NF), and the total time cost. 
For the GB, both initial-value searching and complete-data optimization contribute to the reported NF and time cost. 
As described in Section~\ref{Sec:inv}, the computational cost for sensitivity computation is assumed to be twice that of forward predictions and is included in the NF count. 
Since GF groups rely solely on forward predictions, NF serves as a reliable metric for assessing the computational effort required for parameter identification.

We observe that GB yields the minimal objective function value and demonstrates superior adaptability across various charging currents. 
Although the RMSE at $0.2\rm~C$ is slightly higher than that of GF(I), the RMSE values for other conditions are significantly lower than those of the GF groups. 
In terms of computational cost, GB requires only 250 forward predictions, representing a reduction of over $96\%$ compared to the GF groups, which demand $7424$, $9984$, and $12032$ predictions, respectively. 
Furthermore, GB achieves a $72\%$ decrease in computational time compared to the fastest GF group.
This high efficiency stems from the use of sensitivity information provided by \texttt{DiffLiB}, which guides the optimization process and avoids time-consuming stochastic search. 
The capability to perform end-to-end sensitivity computations positions \texttt{DiffLiB} as a promising tool for industrial parameter identification applications.

\subsection{3D benchmark problems}
\label{Sec:examples:3d}
While traditional P2D models solely account for electrochemical reactions along the cell thickness, the actual state variable distributions in LIBs exhibit spatial inhomogeneity. 
This phenomenon is particularly pronounced in LIBs for energy storage, where the demand for high energy density drives larger cell dimensions and extended electrode lengths, amplifying spatial variations in state variables.
Accurately quantifying the spatial inhomogeneity is critical for precisely describing the internal reactions of LIBs and expanding the applicability of battery simulation models.
A key feature of \texttt{DiffLiB} is its capability to perform 3D forward predictions using real battery geometries, thereby overcoming the geometric constraints of lower-dimensional models. 
In this subsection, we validate this high-fidelity modeling capacity of \texttt{DiffLiB} by extending the battery simulation models in subsections~\ref{Sec:example:bmk} and \ref{Sec:examples:exp} to 3D.

\begin{figure}[H] \centering
    {\includegraphics[width=0.8\textwidth]{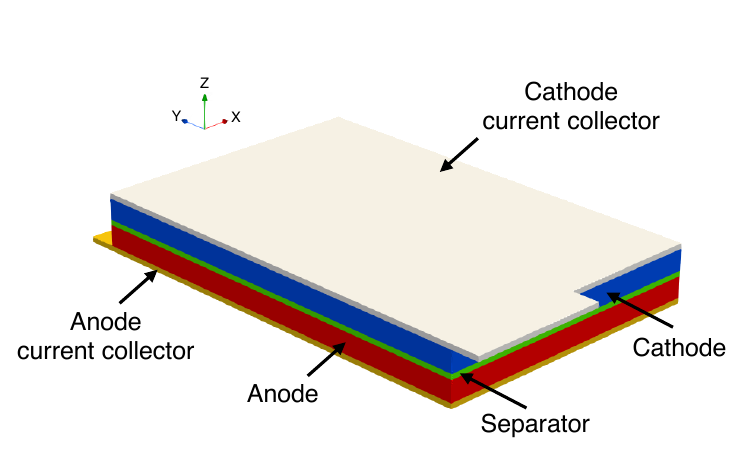}}
    \caption{3D model of the LIB cell from \cite{marquis2019asymptotic}. (Z-axis scaled 100 times.)} \label{fig:3d_model}
\end{figure}

We begin by analyzing the LIB cell from subsection \ref{Sec:example:bmk}. 
Fig.~\ref{fig:3d_model} presents the 3D LIB model with dimensions of $0.137\rm~m$ (width, X), $0.207\rm~m$ (length, Y), and $2.75\cdot10^{-4}\rm~m$ (thickness, Z). To facilitate visualization of the high-aspect-ratio geometry, the Z-axis is scaled by a factor of 100. The anode and cathode current collectors employ copper ($\sigma=5.965\cdot10^7\rm~S/m$) and aluminum ($\sigma=3.55\cdot10^7\rm~S/m$), respectively. The same discharging currents specified in subsection \ref{Sec:example:bmk} are applied to the cathode tab, and the anode tab is set to ground.

The macroscopic computational domain is discretized using 656 eight-node hexahedral elements, while each microscopic domain employs 10 linear interval elements. To validate \texttt{DiffLiB}'s accuracy in 3D analysis, we compare the prediction results against \texttt{COMSOL}.

\begin{figure}[H] \centering
    \subfigure[]
    {\includegraphics[width=0.47\textwidth]{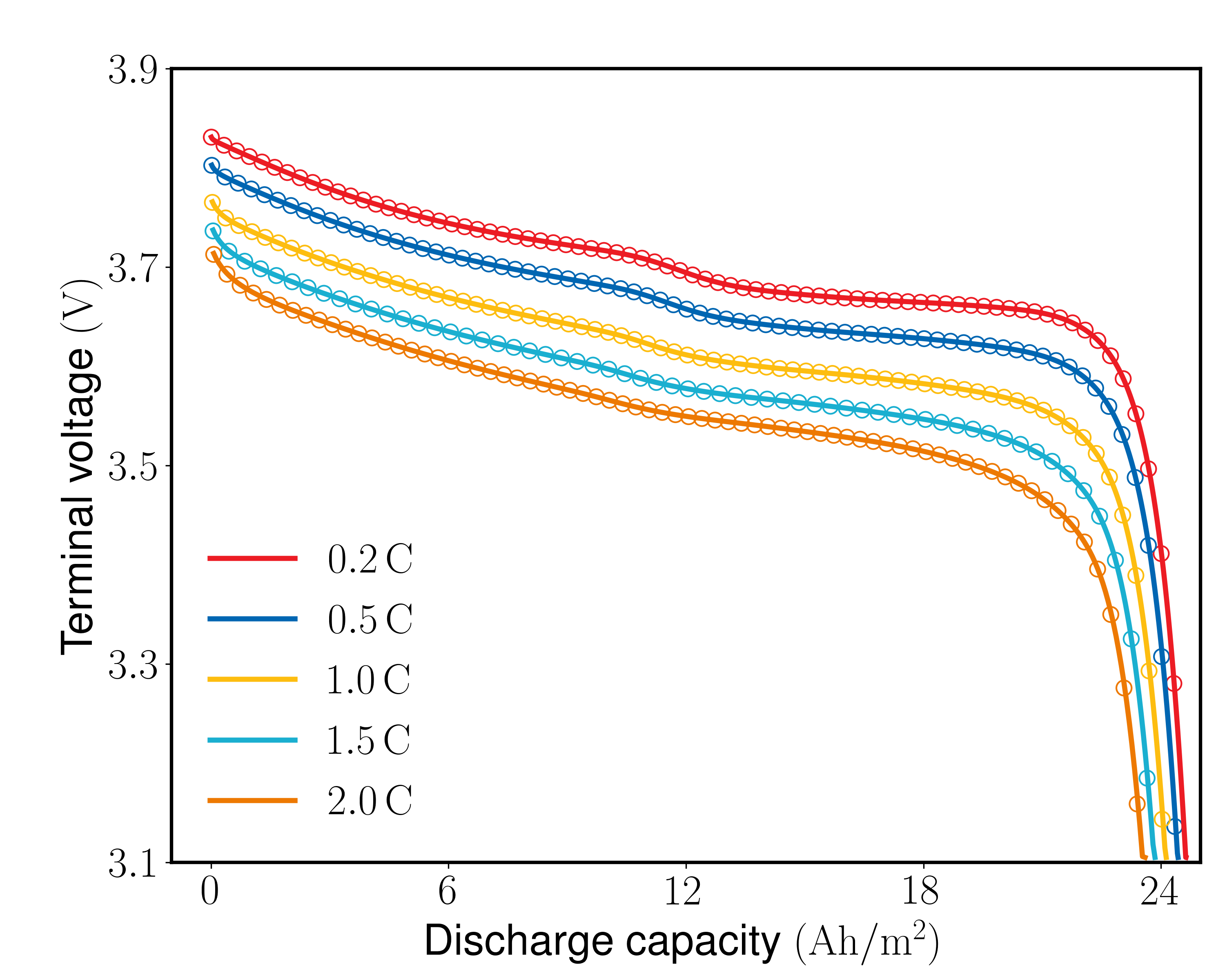}\label{fig:3d_data_a}}
    \hspace{0.01\textwidth}
    \subfigure[]
    {\includegraphics[width=0.47\textwidth]{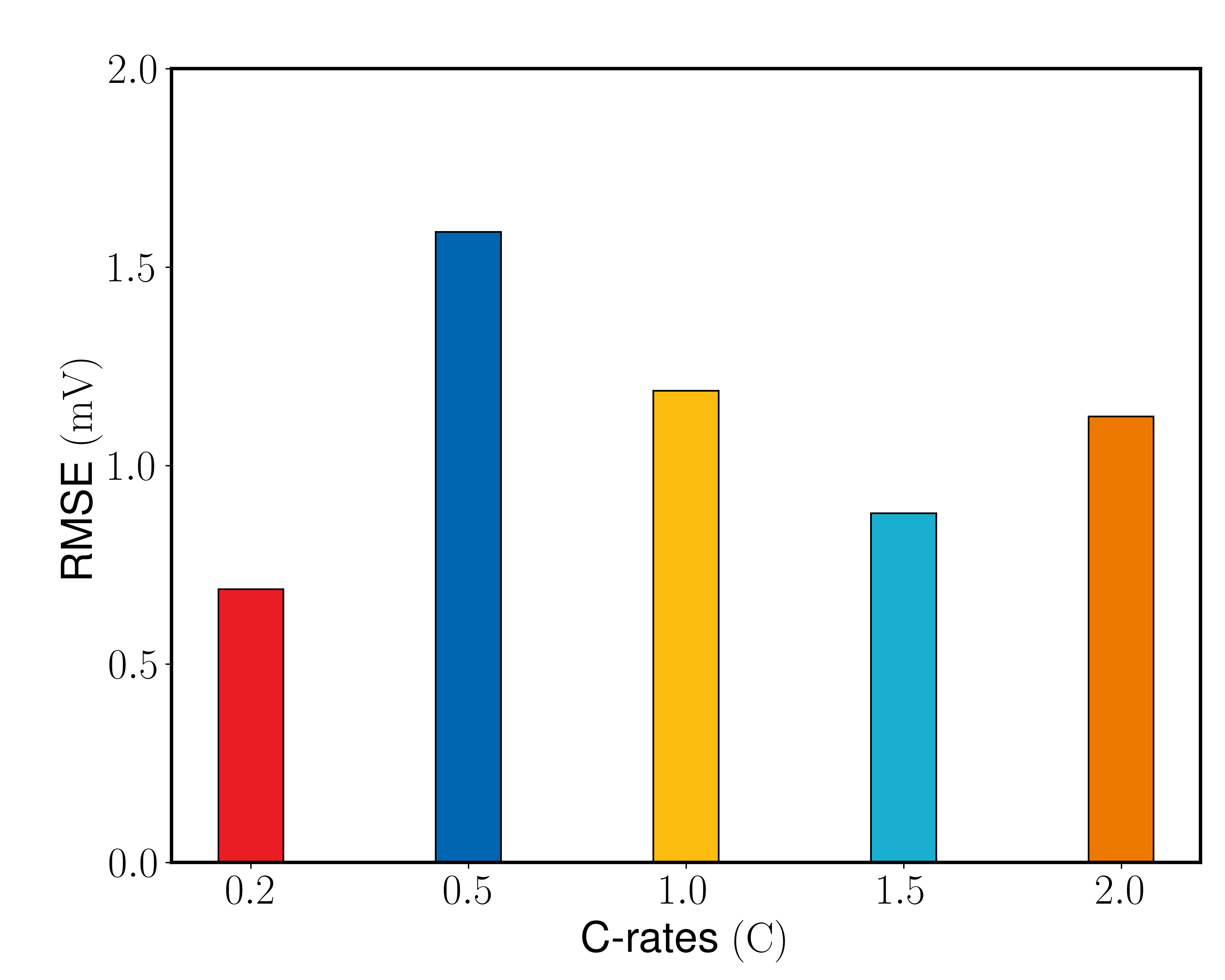}\label{fig:3d_data_b}}
    \caption{3D validation of the terminal voltage under different discharging currents. (Solid lines: \texttt{DiffLiB}; markers: \texttt{COMSOL}). (a) Variation of the terminal voltage during the discharging process. (b) Root mean square errors (RMSE) of the terminal voltage. } \label{fig:3d_data}
\end{figure}

Fig.~\ref{fig:3d_data_a} illustrates the predicted terminal voltage under various discharging currents for the 3D LIB model, with the corresponding RMSE between \texttt{DiffLiB} and \texttt{COMSOL} presented in Fig.~\ref{fig:3d_data_b}. The predictions from \texttt{DiffLiB} for the 3D LIB model show strong agreement with those from \texttt{COMSOL}. Although some RMSE values are slightly higher than those reported in Section~\ref{Sec:example:bmk}, the maximum RMSE remains below $2\rm~mV$, representing less than $1\%$ of the predicted terminal voltage. This indicates the high accuracy of \texttt{DiffLiB} for 3D forward predictions.

\begin{figure}[H] \centering
    \subfigure[]
    {\includegraphics[width=0.9\textwidth]{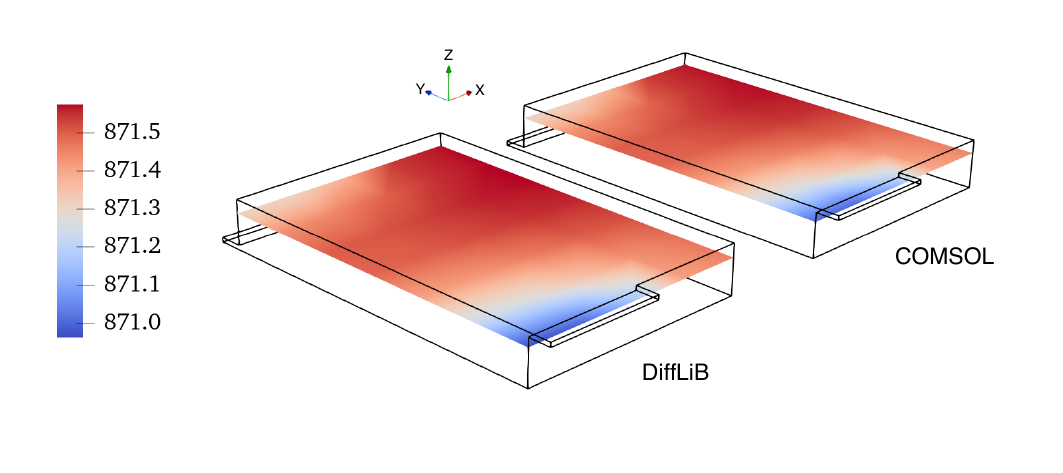}}
    \\
    \subfigure[]
    {\includegraphics[width=0.9\textwidth]{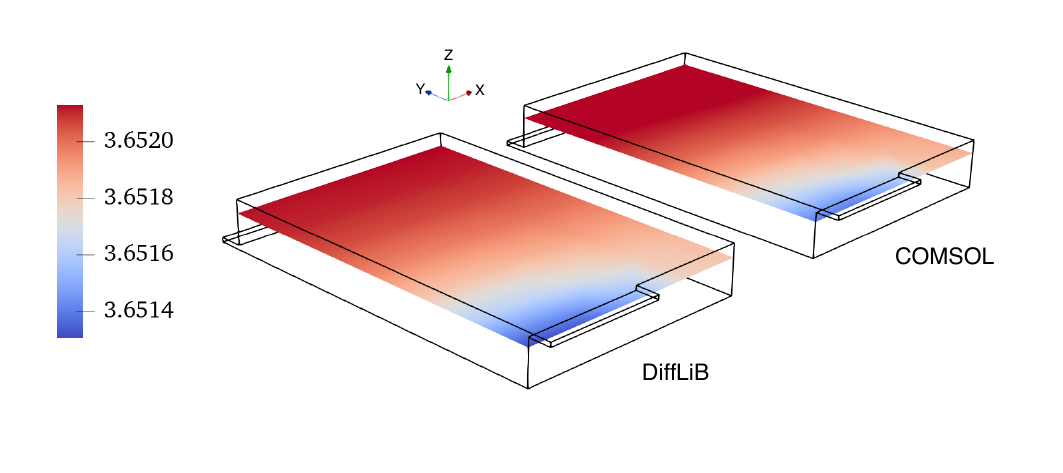}}
    \caption{Spatial distribution of the state variables in the LIB cell from \cite{marquis2019asymptotic}: (a) Electrolyte lithium concentration; (b) Solid-phase potential.} \label{fig:3d_var}
\end{figure}

Fig.~\ref{fig:3d_var} presents the spatial distribution of state variables on the cathode cross-section ($z = 2\cdot10^{-4}\rm~ m$) at $\frac{1}{3}\rm~hrs$ during the $1\rm~C$ discharge, including the electrolyte lithium concentration and solid-phase potential. The results demonstrate well agreement between the \texttt{DiffLiB} and \texttt{COMSOL} predictions, with both models capturing spatial inhomogeneity that are absent in lower-dimensional models.

\begin{figure}[H] \centering
    {\includegraphics[width=0.9\textwidth]{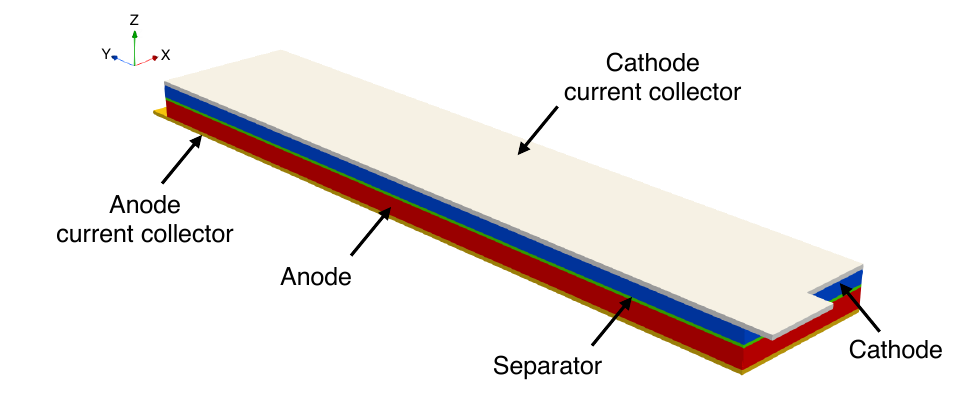}}
    \caption{3D model of the commercial LIB cell. (Z-axis scaled 500 times.)} \label{fig:yang3d_model}
\end{figure}

We next analyze the commercial LIB cell from subsection \ref{Sec:examples:exp}. Fig.~\ref{fig:yang3d_model} shows the corresponding 3D LIB model with dimensions of $0.14855\rm~m$ (width, X) × $0.75\rm~m$ (length, Y) × $8.52\cdot10^{-5}\rm~m$ (thickness, Z), where the Z-axis is scaled by a factor of 500 for visualization clarity. The current collectors maintain the same material and electrical conductivities as the previous 3D model. The model parameters integrate both identified values from subsection~\ref{Sec:examples:exp} and fundamental values determined by the manufacturing information.

The macroscopic computational domain employs $1304$ eight-node hexahedral elements, with each microscopic computational domain discretized using 10 linear interval elements. A $1\rm~C$ charging current is applied at the cathode tab, while the anode tab is set to ground.

\begin{figure}[H] \centering
    \subfigure[]
    {\includegraphics[width=0.9\textwidth]{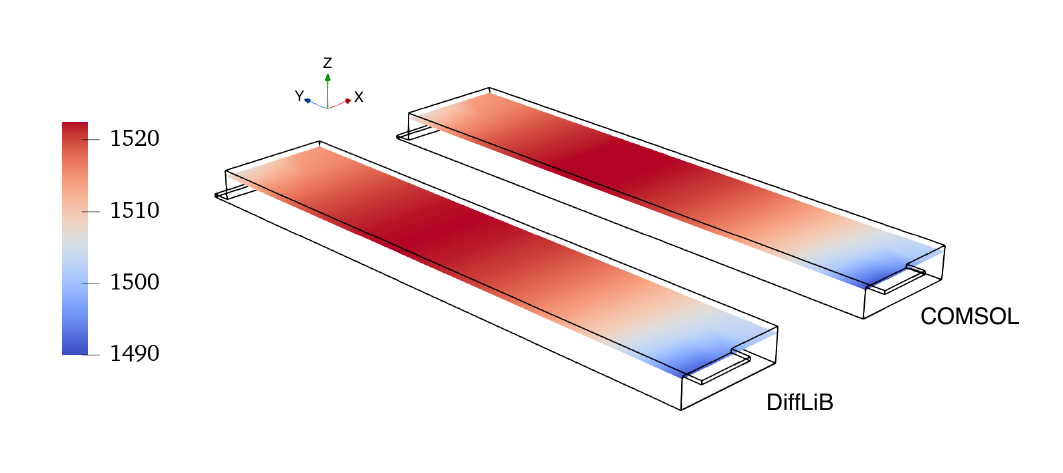}}
    \\
    \subfigure[]
    {\includegraphics[width=0.9\textwidth]{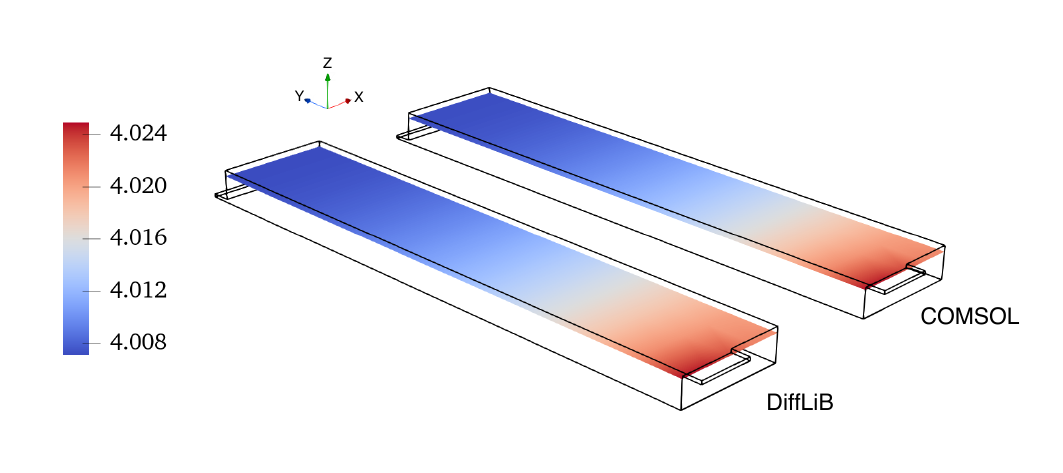}}
    \caption{Spatial distribution of the state variables in the commercial LIB cell: (a) Electrolyte lithium concentration; (b) Solid-phase potential.} \label{fig:yang3d_var}
\end{figure}

Fig.\ref{fig:yang3d_var} shows the distribution of state variables on the cathode cross-section ($z=7\cdot10^{-5}\rm~m$) at $\frac{1}{2}\rm hrs$, including the electrolyte lithium concentration and the solid-phase potential. 
We observe that \texttt{DiffLiB} shows well agreement with \texttt{COMSOL} in the variable distributions again. 
The elongated Y dimension of this LIB cell induces more pronounced spatial inhomogeneity, exhibiting a maximum electrolyte lithium concentration difference of $32\rm~mol/m^3$ across the section.
This high-fidelity modeling capability of \texttt{DiffLiB} is particularly valuable for developing complex-geometry and high-performance LIBs, as it can accurately reflect spatial distributions to characterize internal states and guide design optimization.

\subsection{Parameter identification with 3D models}
\label{Sec:examples:3d_inv}
The preceding examples illustrate that \texttt{DiffLiB} enables efficient gradient-based inverse analysis and high-fidelity modeling with real battery geometries. These capabilities can be seamlessly integrated to address inverse problems with 3D models, facilitating accurate performance prediction and optimization for advanced battery design. Consequently, the final example focuses on a parameter identification problem with the first 3D model introduced in subsection~\ref{Sec:examples:3d}.

The predicted terminal voltage under four discharging currents ($0.5\rm~C$, $1.0\rm~C$, $1.5\rm~C$, and $2.0\rm~C$) serve as the reference data. Seven parameters ($\beta_{\textrm{a}},\,\beta_{\textrm{c}},\,t_+,\,k_{\textrm{s,a}},\,k_{\textrm{s,c}},\,c_{\textrm{s,a}}^{0}\,c_{\textrm{s,c}}^{0}$) are treated as unknowns for identification, with their ranges specified in the second column of Table~\ref{tab:para_3d_inv}. The separator Bruggeman coefficient $\beta_{\textrm{se}}$ is excluded from the optimization as the separator porosity $\varepsilon_{\textrm{se}}=1$ renders it irrelevant to the system response.
The optimization formulation, scaling scheme, and initial-value search strategy (First stage: $0.5\rm~C$; Second stage: $1.0\rm~C$, $1.5\rm~C$, and $2.0\rm~C$) align with those described in subsection~\ref{Sec:examples:exp}. The optimization process terminates when the objective function value falls below $0.001\rm~mV$.

\begin{figure}[H] \centering
    \subfigure[]
    {\includegraphics[width=0.47\textwidth]{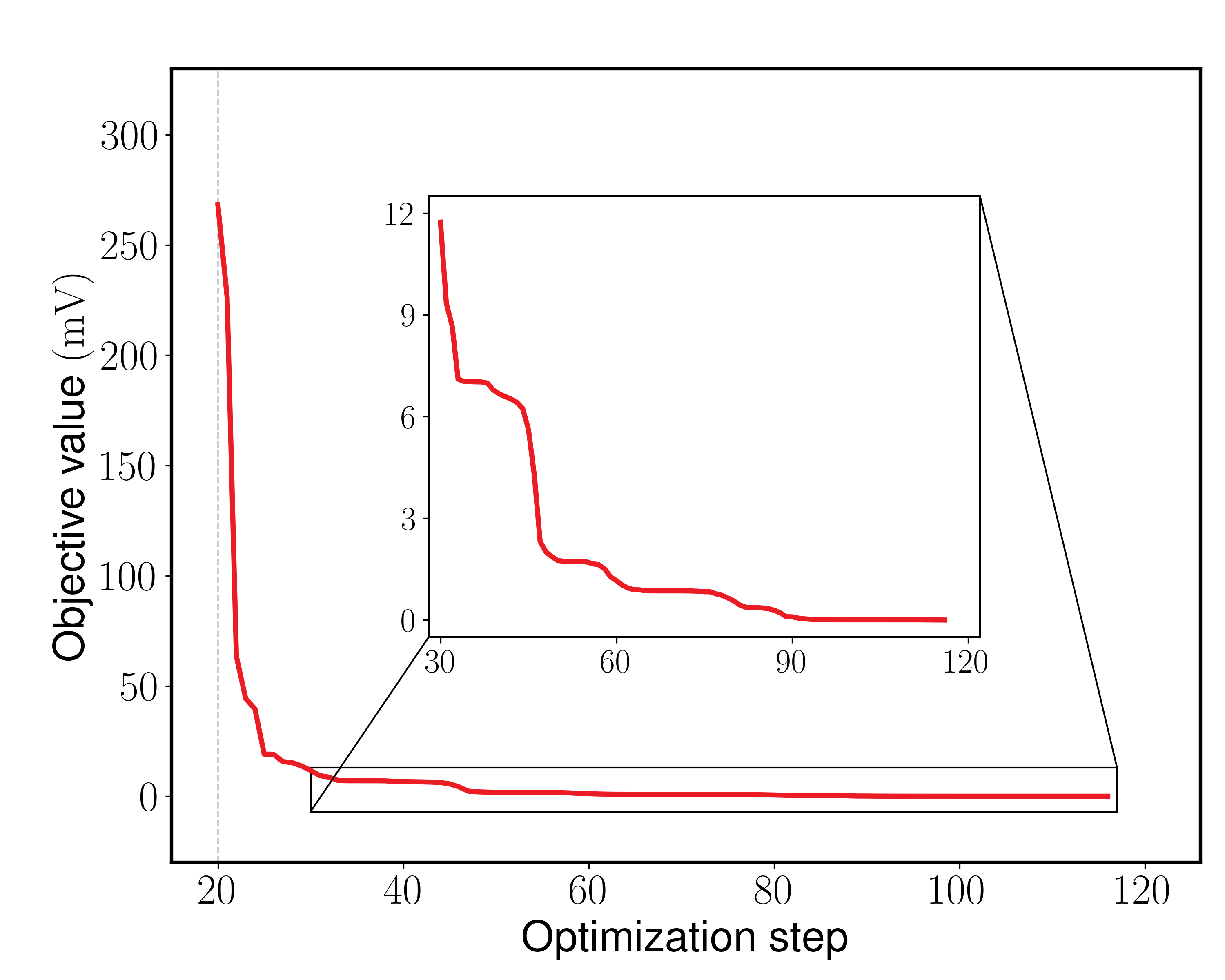}\label{fig:pouch_opt_a}}
    \subfigure[]
    {\includegraphics[width=0.47\textwidth]{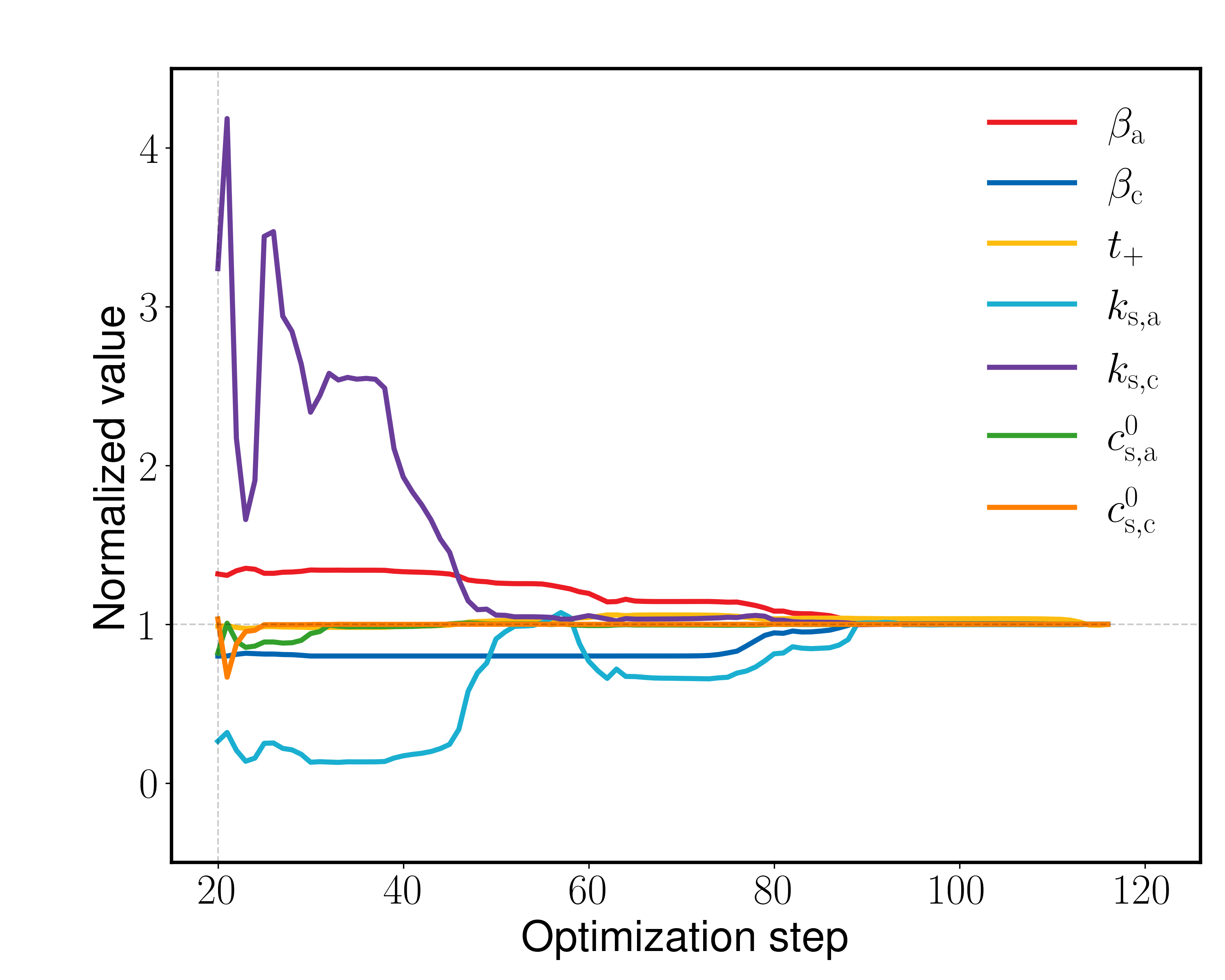}\label{fig:pouch_opt_b}}
    \caption{Optimization history of the parameter identification with 3D models. (a) Objective function value. (b) Normalized parameters.}  \label{fig:pouch_opt}
\end{figure}

Fig.~\ref{fig:pouch_opt} presents the iterations for the complete-data optimization, including the objective function and normalized parameter values. 
As shown in Fig.~\ref{fig:pouch_opt_a}, the objective function value decreases rapidly within the first $10$ iterations before gradually converging to the specified tolerance. 
Fig.~\ref{fig:pouch_opt_b} shows the parameter values normalized by their reference values across the optimization. 
We observe that the reaction rate constants and Bruggeman coefficients for the anode and cathode exhibit significant fluctuations during early iterations. 
This behavior likely stems from complex interdependencies among model parameters. 
For example, the reaction rate constants determine the interfacial reaction through the BV equation (\ref{eq_bv}), thereby affecting multiple reaction processes simultaneously.
All parameters eventually stabilize near their reference values, confirming the robustness of the identification process.

Table~\ref{tab:para_3d_inv} presents the identified parameters alongside their reference values and relative errors. The identified values closely match the reference values with a maximum relative error of $0.15\%$, further validating the effectiveness of the parameter identification scheme. This example demonstrates the capability of \texttt{DiffLiB} to solve inverse problems with 3D models, positioning it as a promising tool for industrial design optimization of LIBs.

\begin{table}[H]\centering
\caption{Comparison of reference and identified model parameters.}
\label{tab:para_3d_inv}
\setlength{\tabcolsep}{4pt} 
\begin{tabularx}{1.0\textwidth}{@{\hspace{5pt}} 
    >{\raggedright\arraybackslash\hsize=0.3\hsize}X 
    >{\centering\arraybackslash\hsize=1.6\hsize}X 
    >{\centering\arraybackslash\hsize=1.0\hsize}X
    *{2}{>{\centering\arraybackslash\hsize=1.0\hsize}X} 
    @{}}
\toprule
Parameter & Range & Reference & Identified & Relative error $\left(\%\right)$ \\
\midrule
$\beta_{\textrm{a}}$  & [1.2, 2.5]  & 1.5 & 1.4994    & 0.0400 \\
$\beta_{\textrm{c}}$  & [1.2, 2.5] & 1.5 & 1.4992    & 0.0533 \\
$t_{+}$  & [0.2, 0.5] & 0.4 & 0.3994    & 0.1500  \\
$k_{\textrm{s,a}}$  & [$5\cdot10^{-12}, 5\cdot10^{-10}$] & $2.0729\cdot10^{-10}$ & $2.0730\cdot10^{-10}$    & 0.0048 \\
$k_{\textrm{s,c}}$  & [$5\cdot10^{-12}, 5\cdot10^{-10}$] & $6.2186\cdot10^{-12}$ & $6.2188\cdot10^{-12}$    & 0.0032 \\
$c_{\textrm{s,a}}^0$ & [$0.6\cdot c_{\textrm{s,a}}^{\textrm{max}}, 0.9\cdot c_{\textrm{s,a}}^{\textrm{max}}$]  & $19987$ & $19987.02$    & 0.0001 \\
$c_{\textrm{s,c}}^0$ & [$0.4\cdot c_{\textrm{s,c}}^{\textrm{max}}, 0.7\cdot c_{\textrm{s,c}}^{\textrm{max}}$]  & $30731$ &  $30731.00$ & 0.0001 \\
\bottomrule
\end{tabularx}
\end{table}

\section{Conclusions and future work}
\label{Sec:conclusion}
In this study, we introduce a differentiable simulation framework \texttt{DiffLiB} for solving forward and inverse problems of LIBs. This framework implements the full-order DFN model through multi-scale finite element formulations, enabling high-fidelity modeling with real battery geometries. Leveraging the AD technique, the Jacobian matrix in the forward problem can be automatically computed, eliminating the need for manual derivations. The customized VJP rule with the implicit differentiation method enables the automatic computation of sensitivity information for the input parameters, facilitating efficient solution of inverse problems. Four numerical examples, encompassing both forward predictions and inverse parameter identification in 2D and 3D, are presented to validate the accuracy and efficiency of \texttt{DiffLiB}. The benchmarking results against \texttt{COMSOL} demonstrate that \texttt{DiffLiB} accurately captures the system response and quantify spatial inhomogeneities through 3D geometry modeling. For the parameter identification task with experimentally measured voltage, the sensitivity information obtained via the customized AD effectively guides the optimization process, reducing forward predictions by $96\%$ and total computation time by $72\%$ compared to gradient-free methods. The 3D parameter identification example further demonstrates \texttt{DiffLiB}'s effectiveness in solving inverse problems with real battery geometries. These features render \texttt{DiffLiB} as a promising tool for the development of complex-geometry and high-performance LIBs.

Future work could focus on incorporating additional physical phenomena, such as thermal-mechanical effects and aging mechanisms. This would facilitate a more precise description of the reactions within real batteries. In addition to C-rate tests, other experiments such as cycling and pulse charging tests can also be considered to broaden the range of applications. Moreover, the inverse problems in different scenarios could be explored, such as the optimization of geometry and manufacturing parameters.

\section{Acknowledgments}

We acknowledge the support from the Hong Kong Innovation and Technology Support Program ITS/032/23 and Beijing DP Technology Co., Ltd.




\bibliographystyle{unsrt}
\bibliography{refs}

\appendix 

\titleformat{\section}[block] 
  {\normalfont\Large\bfseries} 
  {} 
  {0pt} 
  {}
  [] 

\renewcommand\theequation{\thesection.\arabic{equation}} 
\renewcommand\thefigure{\thesection.\arabic{figure}} 
\renewcommand\thetable{\thesection.\arabic{table}} 

\section{Appendix A. Adjoint sensitivity analysis}
\label{app_adj}

\setcounter{equation}{0} 
\setcounter{figure}{0} 
\setcounter{table}{0} 
Consider the objective function with multi-time responses, we first transfer the optimization problem (\ref{eq_inv}) into an unconstrained problem by constructing the following Lagrangian function:
\begin{align}\label{eq_lag}
    \widehat{\mathcal{L}} = \mathcal{L} + \sum_{i=1}^N\left(\boldsymbol{\lambda}^i\right)^{\textrm{T}}\boldsymbol{R}^i+\left(\boldsymbol{\lambda}^0\right)^{\textrm{T}}\left[\boldsymbol{g}\left(\boldsymbol{\theta}\right)-\boldsymbol{U}^0\right]
\end{align}
where $\boldsymbol{\lambda}$ denotes the Lagrangian multiplier. The necessary condition for the optimality of $\widehat{\mathcal{L}}$ can be obtained by seeking its stationary points, according to which the derivative of $\widehat{\mathcal{L}}$ with respect to $\boldsymbol{\lambda}$ and $\boldsymbol{U}$ should satisfy:
\begin{align}\label{eq_dLdl0}
    \frac{\partial\widehat{\mathcal{L}}}{\partial\boldsymbol{\boldsymbol{\lambda}}^0} = \boldsymbol{g}\left(\boldsymbol{\theta}\right)-\boldsymbol{U}^0=\boldsymbol{0}
\end{align}
\begin{align}\label{eq_dLdl}
    \frac{\partial\widehat{\mathcal{L}}}{\partial\boldsymbol{\boldsymbol{\lambda}}^k} = \boldsymbol{R}^k=\boldsymbol{0},\quad k=1,\cdots,N
\end{align}
\begin{align}\label{eq_dLdU0}
    \frac{\partial\widehat{\mathcal{L}}}{\partial\boldsymbol{\boldsymbol{U}}^0} = \left(\boldsymbol{\lambda}^1\right)^{\textrm{T}}\frac{\partial\boldsymbol{R}^1}{\partial\boldsymbol{U}^0}-\left(\boldsymbol{\lambda}^0\right)^{\textrm{T}}=\boldsymbol{0}
\end{align}
\begin{align}\label{eq_dLdU}
    \frac{\partial\widehat{\mathcal{L}}}{\partial\boldsymbol{\boldsymbol{U}}^k} = \frac{\partial\mathcal{L}}{\partial\boldsymbol{\boldsymbol{U}}^k} + \sum_{i=1}^N\left(\boldsymbol{\lambda}^i\right)^{\textrm{T}}\frac{\partial\boldsymbol{R}^i}{\partial\boldsymbol{U}^k}=\boldsymbol{0},\quad k=1,\cdots,N
\end{align}
where (\ref{eq_dLdl0}) corresponds to the initial condition, and (\ref{eq_dLdl}) is the equilibrium equation at each time step. The derivative of $\widehat{\mathcal{L}}$ with respect to $\boldsymbol{\theta}$ can be expressed as:
\begin{align}\label{eq_dLdt}
     \frac{\partial\widehat{\mathcal{L}}}{\partial\boldsymbol{\theta}} = \frac{\partial\mathcal{L}}{\partial\boldsymbol{\theta}} + \sum_{i=1}^N\left(\boldsymbol{\lambda}^i\right)^{\textrm{T}}\frac{\partial\boldsymbol{R}^i}{\partial\boldsymbol{\theta}}+\left(\boldsymbol{\lambda}^0\right)^{\textrm{T}}\frac{\textrm{d}\boldsymbol{g}}{\textrm{d}\boldsymbol{\theta}}
\end{align}

Then we will prove that if (\ref{eq_dLdl0}) and (\ref{eq_dLdl}) are satisfied, and the Lagrange multipliers are obtained through (\ref{eq_dLdU0}) and (\ref{eq_dLdU}), the total derivative $\frac{\textrm{d}\mathcal{L}}{\textrm{d}\boldsymbol{\theta}}$ is equal to (\ref{eq_dLdt}). Substitute (\ref{eq_dLdU}) into (\ref{eq_inv_chain_multi}), we have:
\begin{align}\label{eq_dLdt_simp1}
    \frac{\textrm{d}\mathcal{L}}{\textrm{d}\boldsymbol{\theta}} = \frac{\partial\mathcal{L}}{\partial\boldsymbol{\theta}} - \sum_{i=1}^N\left(\boldsymbol{\lambda}^i\right)^{\textrm{T}}\left(\sum_{k=1}^{N}\frac{\partial\boldsymbol{R}^i}{\partial\boldsymbol{U}^k}\frac{\textrm{d}\boldsymbol{U}^k}{\textrm{d}\boldsymbol{\theta}}\right)
\end{align}
Then we take the derivative of the equilibrium equation $\boldsymbol{R}^i=\boldsymbol{0}$ with respect to $\boldsymbol{\theta}$:
\begin{align}\label{eq_dR}
    \frac{\textrm{d}\boldsymbol{R}^i}{\textrm{d}\boldsymbol{\theta}} = \frac{\partial\boldsymbol{R}^i}{\partial\boldsymbol{\theta}} + \sum_{k=1}^N\frac{\partial\boldsymbol{R}^i}{\partial\boldsymbol{U}^k}\frac{\textrm{d}\boldsymbol{U}^k}{\textrm{d}\boldsymbol{\theta}} + \frac{\partial\boldsymbol{R}^i}{\partial\boldsymbol{U}^0}\frac{\textrm{d}\boldsymbol{g}}{\textrm{d}\boldsymbol{\theta}} =\boldsymbol{0},\quad i=1,\cdots,N
\end{align}
Substitute (\ref{eq_dR}) into (\ref{eq_dLdt_simp1}), we have:
\begin{align}\label{eq_dLdt_simp2}
     \frac{\textrm{d}\mathcal{L}}{\textrm{d}\boldsymbol{\theta}}= \frac{\partial\mathcal{L}}{\partial\boldsymbol{\theta}} + \sum_{i=1}^N\left(\boldsymbol{\lambda}^i\right)^{\textrm{T}}\left[\frac{\partial\boldsymbol{R}^i}{\partial\boldsymbol{\theta}}+\frac{\partial\boldsymbol{R}^i}{\partial\boldsymbol{U}^0}\frac{\textrm{d}\boldsymbol{g}}{\textrm{d}\boldsymbol{\theta}}\right]
\end{align}
Since $\boldsymbol{U}^0$ only contributes to $\boldsymbol{R}^1$, with (\ref{eq_dLdU0}), we obtain: 
\begin{align}\label{eq_dLdt_simp3}
     \frac{\textrm{d}\mathcal{L}}{\textrm{d}\boldsymbol{\theta}}= \frac{\partial\widehat{\mathcal{L}}}{\partial\boldsymbol{\theta}} = \frac{\partial\mathcal{L}}{\partial\boldsymbol{\theta}} + \sum_{i=1}^N\left(\boldsymbol{\lambda}^i\right)^{\textrm{T}}\frac{\partial\boldsymbol{R}^i}{\partial\boldsymbol{\theta}}+\left(\boldsymbol{\lambda}^0\right)^{\textrm{T}}\frac{\textrm{d}\boldsymbol{g}}{\textrm{d}\boldsymbol{\theta}}
\end{align}
which has the same form as (\ref{eq_inv_chain5_single}), with the Lagrangian multipliers corresponding to the adjoint vectors.

The Lagrangian multipliers in (\ref{eq_dLdt_simp3}) can be obtained via (\ref{eq_dLdU0}) and (\ref{eq_dLdU}). Note that $\boldsymbol{U}^N$ only contributes to $\boldsymbol{R}^N$, we can derive the adjoint equation for $\boldsymbol{\lambda}^N$ via (\ref{eq_dLdU}):
\begin{align}\label{eq_ad_it_N}
    \left(\frac{\partial\boldsymbol{R}^N}{\partial\boldsymbol{U}^N}\right)^{\textrm{T}}\boldsymbol{\lambda}^N = -\left(\frac{\partial\mathcal{L}}{\partial\boldsymbol{\boldsymbol{U}}^N}\right)^{\textrm{T}}
\end{align}
For $i=N-1$ to $i=1$, $\boldsymbol{U}^{i}$ contributes to $\boldsymbol{R}^i$ and $\boldsymbol{R}^{i+1}$, with which (\ref{eq_dLdU}) can be simplified as:
\begin{align}\label{eq_dLdU_simp}
    \frac{\partial\mathcal{L}}{\partial\boldsymbol{\boldsymbol{U}}^i} + \left(\boldsymbol{\lambda}^i\right)^{\textrm{T}}\frac{\partial\boldsymbol{R}^i}{\partial\boldsymbol{U}^i} + \left(\boldsymbol{\lambda}^{i+1}\right)^{\textrm{T}}\frac{\partial\boldsymbol{R}^{i+1}}{\partial\boldsymbol{U}^{i}}=0,\quad i=N-1,\cdots,1 
\end{align}
Then the corresponding Lagrangian multipliers can be solved in a backward recurrence with the following adjoint equation:
\begin{align}\label{eq_ad_it}
     \left(\frac{\partial\boldsymbol{R}^i}{\partial\boldsymbol{U}^i}\right)^{\textrm{T}}\boldsymbol{\lambda}^i =-\left[\frac{\partial\mathcal{L}}{\partial\boldsymbol{\boldsymbol{U}}^i}+\left(\boldsymbol{\lambda}^{i+1}\right)^{\textrm{T}}\frac{\partial\boldsymbol{R}^{i+1}}{\partial\boldsymbol{U}^{i}}\right]^{\textrm{T}},\quad i=N-1,\cdots,1 
\end{align}
Finally, $\boldsymbol{\lambda}^0$ can be obtained via (\ref{eq_dLdU0}):
\begin{align}\label{eq_ad_it_0}
    \boldsymbol{\lambda}^0 = \left(\frac{\partial\boldsymbol{R}^1}{\partial\boldsymbol{U}^0}\right)^{\textrm{T}}\boldsymbol{\lambda}^1
\end{align}
We observe that the derivation yields the same expression for $\frac{\partial\mathcal{L}}{\partial\boldsymbol{\theta}}$, and the adjoint equations are consistent with those in the customized reverse-mode AD, further confirming the validity of the AD-based sensitivity analysis framework.

\section{Appendix B. Model parameters and functions I}
\label{app_para_Ma2019}

\setcounter{equation}{0} 
\setcounter{figure}{0} 
\setcounter{table}{0} 

The model parameters and functions for subsection~\ref{Sec:example:bmk} are taken from \cite{marquis2019asymptotic}. The electrolyte diffusivity is expressed as:
\begin{align}
    D_{\textrm{e}} = 5.34\cdot 10^{-10}\cdot\exp\left(-0.65\cdot\frac{c_{\textrm{e}}}{1000}\right)
\end{align}
The electrolyte conductivity is expressed as:
\begin{align}
    \kappa_{\textrm{e}} = 0.0911+1.9101\cdot\left(\frac{c_\textrm{e}}{1000}\right)-1.0520\cdot\left(\frac{c_{\textrm{e}}}{1000}\right)^2+0.1554\cdot\left(\frac{c_\textrm{e}}{1000}\right)^3
\end{align}
The rest parameters are given in Table~\ref{tab:para_bmk}.
\begin{table}[H]\centering
\caption{Model parameters taken from \cite{marquis2019asymptotic}.}
\label{tab:para_bmk}
\setlength{\tabcolsep}{6pt} 
\begin{tabularx}{1.0\textwidth}{
  >{\raggedright}p{28pt}  
  c @{\hspace{\dimexpr\tabcolsep + 5pt}} 
  >{\raggedright\arraybackslash}p{142.7pt} 
  @{\hspace{\dimexpr\tabcolsep - 15pt}}
  c
  @{\hspace{\dimexpr\tabcolsep + 10pt}}
  c
  @{\hspace{\dimexpr\tabcolsep + 10pt}}
  c
}
\toprule
Parameter & Unit & Description & Anode & Separator & Cathode \\
\midrule
$L$ & m & Thickness & $100\cdot10^{-6}$ & $25\cdot10^{-6}$ & $100\cdot10^{-6}$ \\
$\beta$ & $-$ & Bruggeman coefficient & 1.5 & 1.5 & 1.5 \\
$\varepsilon_\textrm{e}$ & $-$ & Electrolyte volume fraction & 0.3 & 1.0 & 0.3 \\
$\varepsilon_\textrm{s}$ & $-$ & Active material volume fraction & 0.6 & $-$ & 0.5 \\
$\varepsilon_\textrm{f}$ & $-$ & Filler volume fraction & 0.1 & $-$ & 0.2 \\
$r_{\textrm{s}}$ & m & Particle radius & $10\cdot10^{-6}$ & $-$ & $10\cdot10^{-6}$ \\
$c^0_{\textrm{s}}$ & $\textrm{mol}/\textrm{m}^3$ & Initial microscopic lithium concentration & 19987 & $-$ & 30731 \\
$c_{\textrm{s}}^{\textrm{max}}$ & $\textrm{mol}/\textrm{m}^3$ & Maximum microscopic lithium concentration & 24983 & $-$ & 51218 \\
$D_{\textrm{s}}$ & $\textrm{m}^2/\textrm{s}$ & Microscopic diffusivity & $3.9\cdot10^{-14}$ & $-$ & $1\cdot10^{-13}$ \\
$k_{\textrm{s}}$ & $\textrm{m}^{2.5}/(\textrm{mol}^{0.5}~\textrm{s})$ & Reaction rate constant & $2.0729\cdot10^{-10}$ & $-$ & $6.2186\cdot10^{-12}$ \\
$\sigma$ & $\textrm{S}/\textrm{m}$ & Electrode conductivity & 100 & $-$ & 10 \\
$c_\textrm{e}^0$ & $\textrm{mol}/\textrm{m}^3$ & Initial electrolyte lithium concentration &  & 1000 &  \\
$t_{+}$ & $-$ & Transference number &  & 0.4 &  \\
$R$ & $\textrm{J}/(\textrm{mol}~\textrm{K})$ & Gas constant &  & 8.314 &  \\
$T$ & $\textrm{K}$ & Temperature &  & 298.15 &  \\
$F$ & $\textrm{C}/\textrm{mol}$ & Faraday constant &  & 96485 &  \\
\bottomrule
\end{tabularx}
\end{table}

The OCP function of the anode is expressed as:
\begin{align}
    U_{\textrm{oc,a}}=\ &0.194 + 1.5\cdot\exp\left(-120\cdot\theta_{\mathrm{a}}\right)+0.0351\cdot\tanh\left(\frac{\theta_{\mathrm{a}}-0.286}{0.083}\right)\nonumber\\
    -&0.0045\cdot\tanh\left(\frac{\theta_{\mathrm{a}}-0.849}{0.119}\right)-0.035\cdot\tanh\left(\frac{\theta_{\textrm{a}}-0.9233}{0.05}\right)\nonumber\\
    -&0.0147\cdot\tanh\left(\frac{\theta_{\textrm{a}}-0.5}{0.034}\right)-0.102\cdot\tanh\left(\frac{\theta_{\textrm{a}}-0.194}{0.142}\right)\nonumber\\
    -&0.022\cdot\tanh\left(\frac{\theta_{\textrm{a}}-0.9}{0.0164}\right)-0.011\cdot\tanh\left(\frac{\theta_{\textrm{a}}-0.124}{0.0226}\right)\nonumber\\
    +&0.0155\cdot\tanh\left(\frac{\theta_{\textrm{a}}-0.105}{0.029}\right)
\end{align}
where $\theta_{\textrm{a}} = \frac{c_{\textrm{s,a}}^{\textrm{surf}}}{c_{\textrm{s,a}}^{\textrm{max}}}$. The OCP function of the cathode is expressed as:
\begin{align}
        U_{\textrm{oc,c}}=\ &0.07645\cdot\tanh\left(30.834-54.4806\cdot\theta_{\textrm{c}}\right)
        +2.1581\cdot\tanh\left(52.294-50.294\cdot\theta_{\textrm{c}}\right)\nonumber\\
        -&0.14169\cdot\tanh\left(11.0923-19.8543\cdot\theta_{\textrm{c}}\right)+0.2051\cdot\tanh\left(1.4684-5.4888\cdot\theta_{\textrm{c}}\right)\nonumber\\
        +&0.2531\cdot\tanh\left(\frac{0.56478-\theta_{\textrm{c}}}{0.1316}\right)-0.02167\cdot\tanh\left(\frac{\theta_{\textrm{c}}-0.525}{0.006}\right)+2.16216
\end{align}
where $\theta_{\textrm{c}} = 1.062\cdot \frac{c_{\textrm{s,c}}^{\textrm{surf}}}{c_{\textrm{s,c}}^{\textrm{max}}}$.

\section{Appendix C. Model parameters and functions II}
\label{app_exp_para}

\setcounter{equation}{0} 
\setcounter{figure}{0} 
\setcounter{table}{0} 

For the commercial $\rm LiCoO_2$-Graphite battery cell in subsection~\ref{Sec:examples:exp}, some parameters and functions have been determined by the manufacturing information. The electrolyte conductivity is expressed as:
\begin{align}
    \kappa_{\textrm{e}} = 2.915\cdot\left(\frac{c_{\textrm{e}}}{1000}\right)-2.238\cdot\left(\frac{c_{\textrm{e}}}{1000}\right)^{1.5}+0.1147\cdot\left(\frac{c_{\textrm{e}}}{1000}\right)^3
\end{align}
The rest parameters are given in Table~\ref{tab:para_exp}, where the unknown parameters are denoted as TBI (to be identified). 

\begin{table}[H]\centering
\caption{Model parameters of a commercial LIB cell. (TBI: to be identified.)}
\label{tab:para_exp}
\setlength{\tabcolsep}{6pt} 
\begin{tabularx}{1.0\textwidth}{
  >{\raggedright}p{28pt}  
  c @{\hspace{\dimexpr\tabcolsep + 5pt}} 
  >{\raggedright\arraybackslash}p{142.7pt} 
  @{\hspace{\dimexpr\tabcolsep - 15pt}}
  c
  @{\hspace{\dimexpr\tabcolsep + 10pt}}
  c
  @{\hspace{\dimexpr\tabcolsep + 10pt}}
  c
}
\toprule
Parameter & Unit & Description & Anode & Separator & Cathode \\
\midrule
$L$ & m & Thickness & $36.71\cdot10^{-6}$ & $5.40\cdot10^{-6}$ & $26.13\cdot10^{-6}$ \\
$\beta$ & $-$ & Bruggeman coefficient & $\textrm{TBI}$ & $\textrm{TBI}$ & $\textrm{TBI}$ \\
$\varepsilon_\textrm{e}$ & $-$ & Electrolyte volume fraction & 0.32533  & 0.40000 & 0.21542 \\
$\varepsilon_\textrm{s}$ & $-$ & Active material volume fraction & 0.65202  & $-$ & 0.76142 \\
$\varepsilon_\textrm{f}$ & $-$ & Filler volume fraction & 0.02265 & $-$ & 0.02316 \\
$r_{\textrm{s}}$ & m & Particle radius & $11.4\cdot10^{-6}$ & $-$ &  $7.5\cdot10^{-6}$\\
$c^0_{\textrm{s}}$ & $\textrm{mol}/\textrm{m}^3$ & Initial microscopic lithium concentration & $\textrm{TBI}$ & $-$ & $\textrm{TBI}$ \\
$c_{\textrm{s}}^{\textrm{max}}$ & $\textrm{mol}/\textrm{m}^3$ & Maximum microscopic lithium concentration & 28408 & $-$ & 51568 \\
$D_{\textrm{s}}$ & $\textrm{m}^2/\textrm{s}$ & Microscopic diffusivity & $1.3579\cdot10^{-10}$ & $-$ & $2.4700\cdot10^{-12}$ \\
$k_{\textrm{s}}$ & $\textrm{m}^{2.5}/(\textrm{mol}^{0.5}~\textrm{s})$ & Reaction rate constant & $\textrm{TBI}$ & $-$ & $\textrm{TBI}$ \\
$\sigma$ & $\textrm{S}/\textrm{m}$ & Electrode conductivity & 100 & $-$ & 10 \\
$D_{\textrm{e}}$ & $\textrm{m}^2/\textrm{s}$ & Electrolyte diffusivity  &  & $1.93\cdot10^{-10}$ &  \\
$c_\textrm{e}^0$ & $\textrm{mol}/\textrm{m}^3$ & Initial electrolyte lithium concentration &  & 1150 &  \\
$t_{+}$ & $-$ & Transference number &  & $\textrm{TBI}$ &  \\
$R$ & $\textrm{J}/(\textrm{mol}~\textrm{K})$ & Gas constant &  & 8.314 &  \\
$T$ & $\textrm{K}$ & Temperature &  & 298.15 &  \\
$F$ & $\textrm{C}/\textrm{mol}$ & Faraday constant &  & 96485 &  \\
\bottomrule
\end{tabularx}
\end{table}

The OCP function of the anode is expressed as:
\begin{align}
        U_{\textrm{oc,a}} =\ &2.673022 \cdot \exp[-\left(\frac{\theta_{\textrm{a}} + 0.034828}{0.032734}\right)^2] 
        +0.062721 \cdot \exp[-\left(\frac{\theta_{\textrm{a}} - 0.036085}{0.024854}\right)^2]\nonumber \\
        +\ &0.159337 \cdot \exp[-\left(\frac{\theta_{\textrm{a}} - 0.078792}{0.112443}\right)^2] 
        + 0.064161 \cdot \exp[-\left(\frac{\theta_{\textrm{a}} - 0.302060}{0.187485}\right)^2]\nonumber \\
        +\ &0.019294 \cdot \exp[-\left(\frac{\theta_{\textrm{a}} - 0.487684}{0.068516}\right)^2] 
        +0.081550 \cdot \exp[-\left(\frac{\theta_{\textrm{a}} - 0.747598}{0.671913}\right)^2]
\end{align}
where $\theta_{\textrm{a}} = \frac{c_{\textrm{s,a}}^{\textrm{surf}}}{c_{\textrm{s,a}}^{\textrm{max}}}$. The OCP function of the cathode is expressed as:
\begin{align}
    U_{\textrm{oc,c}} =\ &4.558259 \cdot \exp[-\left(\frac{\theta_{\textrm{c}} - 0.154080}{0.748906}\right)^2] 
    +1.561895 \cdot \exp[-\left(\frac{\theta_{\textrm{c}} - 0.861132}{0.328293}\right)^2]\nonumber \\
    +\ &0.058271 \cdot \exp[-\left(\frac{\theta_{\textrm{c}} - 0.888260}{0.024912}\right)^2] 
    +0.620818 \cdot \exp[-\left(\frac{\theta_{\textrm{c}} - 0.940724}{0.147502}\right)^2]
\end{align}
where $\theta_{\textrm{c}} = \frac{c_{\textrm{s,c}}^{\textrm{surf}}}{c_{\textrm{s,c}}^{\textrm{max}}}$.

As described in subsection~\ref{Sec:examples:exp}, the unknown parameters are identified through the L-BFGS-B and SEGA, which ranges and identified values are shown in Table~\ref{tab:vars_exp}.

\begin{table}[H]
\centering
\caption{Identified model parameters of L-BFGS-B (GB) and SEGA (GF).}
\label{tab:vars_exp}
\setlength{\tabcolsep}{6pt} 
\begin{tabularx}{\textwidth}{@{\hspace{5pt}}  
    >{\raggedright\arraybackslash\hsize=0.3\hsize}X 
    >{\centering\arraybackslash\hsize=1.7\hsize}X 
    >{\centering\arraybackslash\hsize=1.0\hsize}X
    *{3}{>{\centering\arraybackslash\hsize=1.0\hsize}X} 
@{}}
\toprule
Parameter  & Range  & GB & GF(I)  & GF(II) & GF(III) \\
\midrule
$\beta_{\textrm{a}}$   & [1.5, 2.5] & 2.0573    & 1.5000
    & 1.7987   & 1.5820   \\
$\beta_{\textrm{se}}$   & [1.5, 2.5] & 2.0110    & 2.5000
    & 1.7325   & 1.6691   \\
$\beta_{\textrm{c}}$   & [1.5, 2.5] & 2.5000    & 2.1896    & 2.2227   & 1.9475  \\
$t_{+}$   & [0.2, 0.5] & 0.3261    & 0.2069    & 0.4461   & 0.3932   \\
$k_{\textrm{s,a}}$   & [$5\cdot10^{-12}, 5\cdot10^{-10}$] & $2.1925\cdot 10^{-11}$    & $1.8355\cdot 10^{-11}$    & $1.5510\cdot 10^{-11}$   & $1.3866\cdot 10^{-11}$   \\
$k_{\textrm{s,c}}$   & [$5\cdot10^{-12}, 5\cdot10^{-10}$]  & $4.7286\cdot10^{-10}$    &$5.0000\cdot10^{-10}$  & $4.1463\cdot 10^{-10}$   & $4.9996\cdot 10^{-10}$   \\
$c_{\textrm{s,a}}^0$   & [$0.001\cdot c_{\textrm{s,a}}^{\textrm{max}}, 0.2\cdot c_{\textrm{s,a}}^{\textrm{max}}$] & $249.33$    & $253.04$    & 258.70   & 263.73   \\
$c_{\textrm{s,c}}^0$   & [$0.8\cdot c_{\textrm{s,c}}^{\textrm{max}}, 0.999\cdot c_{\textrm{s,c}}^{\textrm{max}}$] &  $48145.64$ & $48147.02$    & 48270.62   & 48295.18   \\
\bottomrule
\end{tabularx}
\end{table}

\end{document}